\documentclass[aps,twocolumn,preprintnumbers,superscriptaddress,prl,amsmath,amssymb,nofootinbib,longbibliography]{revtex4-1}

\usepackage{graphicx}
\usepackage{dcolumn}
\usepackage{bm}
\usepackage{color}
\usepackage{paralist}
\usepackage{soul}

\begin{document}

\title{Temperature-induced valence-state transition in double perovskite Ba$_{2-x}$Sr$_{x}$TbIrO$_6$}

\author{Z. Y. Zhao}
\affiliation{Materials Science and Technology Division, Oak Ridge National Laboratory, Oak Ridge, Tennessee 37831, USA}
\affiliation{Department of Physics and Astronomy, University of Tennessee, Knoxville, Tennessee 37996, USA}
\affiliation{State Key Laboratory of Structural Chemistry, Fujian Institute of Research on the Structure of Matter, Chinese Academy of Sciences, Fuzhou, Fujian 350002, People's Republic of China}

\author{S. Calder}
\affiliation{Neutron Scattering Division, Oak Ridge National Laboratory, Oak Ridge, Tennessee 37831, USA}

\author{M. H. Upton}
\affiliation{X-ray Science Division, Argonne National Laboratory, Argonne, Illinois 60439, USA}

\author{H. D. Zhou}
\affiliation{Department of Physics and Astronomy, University of Tennessee, Knoxville, Tennessee 37996, USA}

\author{Z. Z. He}
\affiliation{State Key Laboratory of Structural Chemistry, Fujian Institute of Research on the Structure of Matter, Chinese Academy of Sciences, Fuzhou, Fujian 350002, People's Republic of China}

\author{M. A. McGuire}
\affiliation{Materials Science and Technology Division, Oak Ridge National Laboratory, Oak Ridge, Tennessee 37831, USA}

\author{J.-Q. Yan}
\affiliation{Materials Science and Technology Division, Oak Ridge National Laboratory, Oak Ridge, Tennessee 37831, USA}

\date{\today}

\begin{abstract}

In this work, a temperature-induced valence-state transition is studied in a narrow composition range of Ba$_{2-x}$Sr$_x$TbIrO$_6$ by means of x-ray and neutron powder diffraction, resonant inelastic x-ray scattering, magnetic susceptibility, electrical resistivity, and specific heat measurements. The valence-state transition involves an electron transfer between Tb and Ir leading to the valence-state change between Tb$^{3+}$/Ir$^{5+}$ and Tb$^{4+}$/Ir$^{4+}$ phases. This first-order transition has a dramatic effect on the lattice, transport properties, and the long-range magnetic order at low temperatures for both Tb and Ir ions. Ir$^{5+}$ ion has an electronic configuration of 5$d^4$ ($J\rm_{eff}$ = 0) which is expected to be nonmagnetic. In contrast, Ir$^{4+}$ ion with a configuration of 5$d^5$($J\rm_{eff}$ = 1/2) favors a long-range magnetic order. For $x$ = 0.1 with Tb$^{3+}$/Ir$^{5+}$ configuration to the lowest temperature (2 K) investigated in this work, a spin-glass behavior is observed around 5 K indicating Ir$^{5+}$ ($J\rm_{eff}$ = 0) ions act as a spacer reducing the magnetic interactions between Tb$^{3+}$ ions. For $x$ = 0.5 with Tb$^{4+}$/Ir$^{4+}$ configuration below the highest temperature 400 K of this work, a long-range antiferromagnetic order at $T\rm_N$ = 40 K is observed highlighting the importance of Ir$^{4+}$ ($J\rm_{eff}$ = 1/2) ions in promoting the long-range magnetic order of both Tb and Ir ions. For 0.2 $\leqslant x \leqslant$ 0.375, a temperature-induced valence-state transition from high-temperature Tb$^{3+}$/Ir$^{5+}$ phase to low-temperature Tb$^{4+}$/Ir$^{4+}$ phase occurs in the temperature range 180 K $\leqslant T \leqslant$ 325 K and the transition temperature increases with $x$.  The compositional dependence demonstrates the ability to tune the the valence state for a critical region of $x$ that leads to a concurrent change in magnetism and structure. This tuning ability could be employed with suitable strain in thin films to act as a switch as the magnetism is manipulated.

\end{abstract}

\maketitle

\section{Introduction}

The complex interplay among spin-orbit coupling (SOC), on-site Coulomb interaction, non-cubic crystal field, and electronic bandwidths leads to rich exotic phenomena and novel physics in $4d/5d$ transition-metal compounds \cite{witczak2014correlated,rau2016spin,schaffer2016recent}. One important manifestation is the nontrivial $J\rm_{eff}$ = 1/2 Mott state in Sr$_2$IrO$_4$ with tetravalent Ir$^{4+}$ ($5d^5$) ions, in which the lower $J\rm_{eff}$ = 3/2 band is fully occupied leaving the higher $J\rm_{eff}$ = 1/2 band half filled, see Fig. 1(a) \cite{kim2008novel}. Such spin-orbit-entangled $J\rm_{eff}$ = 1/2 scenario can give rise to unique magnetic ground states in 4$d^5$ and 5$d^5$ compounds, such as Kitaev quantum spin liquid \cite{takagi2019concept}. Compounds containing transition-metal ions with $4d^3$ or $5d^3$ configuration can host magnetism with magnetic ordering temperatures even above room temperature \cite{schnelle2021magnetic,tian2015high, hiley2015antiferromagnetism,  shi2009continuous, rodriguez2011high}. In contrast to the rich magnetic phenomena for $4d/5d$ ions with 3 or 5 electrons, a nonmagnetic state is expected for $J\rm_{eff}$ = 0 compounds in which 5$d$ transition-metal ions have four electrons that are paired in $J\rm_{eff}$ = 3/2 band, see Fig. 1(b). Despite the theoretical prediction of excitonic magnetism and tremendous experimental efforts,  the search for an intrinsic long-range magnetic order in $J\rm_{eff}$ = 0 compounds is still ongoing \cite{khaliullin2013excitonic,meetei2015novel, cao2014novel,nag2016origin, dey2016ba, bhowal2015breakdown, corredor2017iridium, pajskr2016possibility, ranjbar2015structural, chaloupka2016doping, terzic2017evidence, phelan2016influence, fuchs2018unraveling, chen2017magnetism, hammerath2017diluted, zhao2016fragile, wolff2017comparative, prasad2018jeff, laguna2020magnetism, davies2019evidence, gong2018ground}. One challenge is to distinguish the magnetism from lattice defects or unintentional doping.

\begin{figure}
\includegraphics[clip,width=8.5cm]{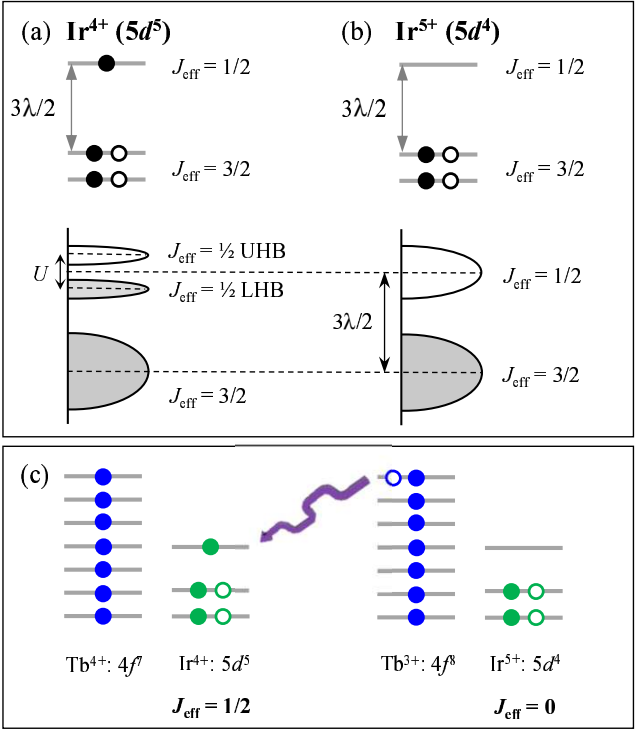}
\caption{(Color online) (a, b) Schematic diagrams of electron filling on $t_{2g}$ level split by spin-orbit coupling (SOC) for Ir$^{4+}$ ($5d^5$) and Ir$^{5+}$ ($5d^4$) configurations. $\lambda$ is the strength of SOC, $U$ is the on-site Coulomb interaction. (c) Electronic configuration of Tb$^{4+}$/Ir$^{4+}$ and Tb$^{3+}$/Ir$^{5+}$. The arrow illustrates the electron transfer from Tb to Ir leading to the valence-state transition.}
\label{diagram}
\end{figure}

The preference of a nonmagnetic ground state for $d^4$ ions indicates possible switch on/off of magnetism by controlling the charge transfer between $d^4$ and $d^3$ or $d^5$ ions. One approach toward this magnetism control is intermetallic charge transfer driven by temperature, chemical or hydrostatic pressure, or other external stimuli such as electric or magnetic fields. The intermetallic charge transfer is quite rare in transition-metal oxides \cite{long2010intermetallic}. However, rare-earth ions Pr, Ce, Tb can stabilize in different oxidization states in perovskite structured oxides which makes possible charge transfer between these rare-earth and transition-metal ions. One good example of such an intermetallic charge transfer is (Pr,Y)$_{1-x}$Ca$_x$CoO$_3$ in which the charge transfer from Pr to Co ions at low temperatures changes the valence state of both Pr and Co ions and induces a metal-insulator transition accompanied with magnetic and structural anomalies \cite{hejtmanek2010metal}.

A similar charge transfer was also reported in Sr$_2$TbRu$_{1-x}$Ir$_x$O$_6$ \cite{zhou2005temperature, doi2000magnetic} and Ba$_2$PrRu$_{1-x}$Ir$_x$O$_6$ \cite{wakeshima2001valence,sannigrahi2019first} where the chemical pressure from mixing Ru and Ir induces a charge transfer between the rare-earth and transition-metal ions. This first-order valence-state transition induces a sudden change of lattice and magnetic properties. The effect of $J\rm_{eff}$ = 0 ions and valence state change on the magnetic properties is well illustrated by the evolution of magnetic properties with Ir substitution in Sr$_2$TbRu$_{1-x}$Ir$_x$O$_6$ \cite{doi2000magnetic}. Sr$_2$Tb$^{3+}$Ru$^{5+}$O$_6$ shows a long-range magnetic order at 41 K. With increasing substitution of Ru$^{5+}$ (4$d^3$) by $J\rm_{eff}$ = 0 Ir$^{5+}$ (5$d^4$) ions, the magnetic order is suppressed and disappeared around $x$ = 0.8. With $x \geqslant$ 0.85, the chemical pressure leads to the transfer of one electron from Tb$^{3+}$ to Ru$^{5+}$/Ir$^{5+}$ site resulting in Sr$_2$Tb$^{4+}$(Ru$_{1-x}$Ir$_x$)$^{4+}$O$_6$ ordered around 50 K. This compositional dependence suggests the nonmagnetic nature of 5$d^4$ Ir$^{5+}$ ($J\rm_{eff}$ = 0) and highlights the importance of 5$d^5$ Ir$^{4+}$ ($J\rm_{eff}$ = 1/2) in the long-range magnetic order of both rare-earth and transition-metal ions in this system.

Mixing Ba and Sr without disturbing the transition-metal site in Ba$_{2-x}$Sr$_x$TbIrO$_6$ was proposed to induce a valence-state transition, which has been studied from the structural point of view at room temperature \cite{zhou2005independent}. Ba$_2$Tb$^{3+}$Ir$^{5+}$O$_6$ ($x$ = 0) crystallizes in a cubic structure with a $Fm\overline{3}m$ space group (see Fig. 2(a)) and paramagnetic down to 2 K \cite{wakeshima2000crystal}, while Sr$_2$Tb$^{4+}$Ir$^{4+}$O$_6$ ($x$ = 2) crystallizes in a monoclinic structure with a $P2_1/n$ space group (see Fig. 2(a)) and undergoes two antiferromagnetic (AFM) transitions at 51 K and 25 K \cite{harada2000magnetic,harada1999structure}. On account of the smaller size of Sr cations, the lower tolerance factor (0.992 for Ba$_2$TbIrO$_6$ \cite{wakeshima2000crystal} and 0.959 for Sr$_2$IrO$_6$ \cite{harada1999structure}) induces the tilting and distortion of the TbO$_6$/IrO$_6$ octahedra and SrO$_{12}$ polyhedra, which is responsible for the change of the crystal structure from cubic to monoclinic. With increasing $x$ in Ba$_{2-x}$Sr$_x$TbIrO$_6$, the lattice parameter shows a sudden drop around $x$ = 0.35 suggesting a valence-state change from Tb$^{3+}$/Ir$^{5+}$ to Tb$^{4+}$/Ir$^{4+}$ phase.

Because the ionic radius of Sr$^{2+}$ is smaller than that of Ba$^{2+}$, partial substitution of Ba$^{2+}$ by Sr$^{2+}$ is expected to reduce the tolerance factor, which measures the stability and distortion of perovskites. The transition from Tb$^{3+}$/Ir$^{5+}$ to Tb$^{4+}$/Ir$^{4+}$ phase with increasing $x$ in Ba$_{2-x}$Sr$_x$TbIrO$_6$ signals that a smaller tolerance factor stabilizes the Tb$^{4+}$/Ir$^{4+}$ phase. For most perovskites, the tolerance factor has a positive temperature dependence. This indicates that Ba$_{2-x}$Sr$_x$TbIrO$_6$ members near the critical composition $x$ = 0.35 can have a temperature-induced valence-state transition between the high-temperature Tb$^{3+}$/Ir$^{5+}$ and the low-temperature Tb$^{4+}$/Ir$^{4+}$ configurations, which hasn't been studied yet. More importantly, the valence-state transition can have a dramatic effect on the magnetic properties considering the change of the electronic configuration of Ir and might be employed to tune magnetism, as shown in Fig. 1(c). In this work, we investigate the thermally-driven valence-state transition in Ba$_{2-x}$Sr$_x$TbIrO$_6$ with the motivation of understanding its effect on the low-temperature magnetism and possible switchable magnetism using the transition between $J\rm_{eff}$ = 0 and $J\rm_{eff}$ = 1/2 at Ir site. The thermally-driven valence-state transition was investigated in the composition range 0.2 $\leqslant x \leqslant$ 0.375 by measuring x-ray and neutron powder diffraction, resonant inelastic x-ray scattering, magnetic susceptibility, electrical resistivity, and specific heat. The valence-state transition has a dramatic effect on the electronic configuration of Ir and thus the long-range magnetic order at low temperatures for both Tb and Ir ions. A magnetic order is observed in Ba$_{2-x}$Sr$_x$Tb$^{4+}$Ir$^{4+}$O$_6$ but absent in Ba$_{2-x}$Sr$_x$Tb$^{3+}$Ir$^{5+}$O$_6$. The valence-state change is associated with a dramatic first-order structural change of the lattice. Consequently, control of the lattice, through strain or pressure, is a potential route to further control and drive the valence state and associated magnetic order. This would yield a switchable mechanism to turn on and off the magnetism which deserves further investigation.

\begin{figure*}
\includegraphics[width=15cm]{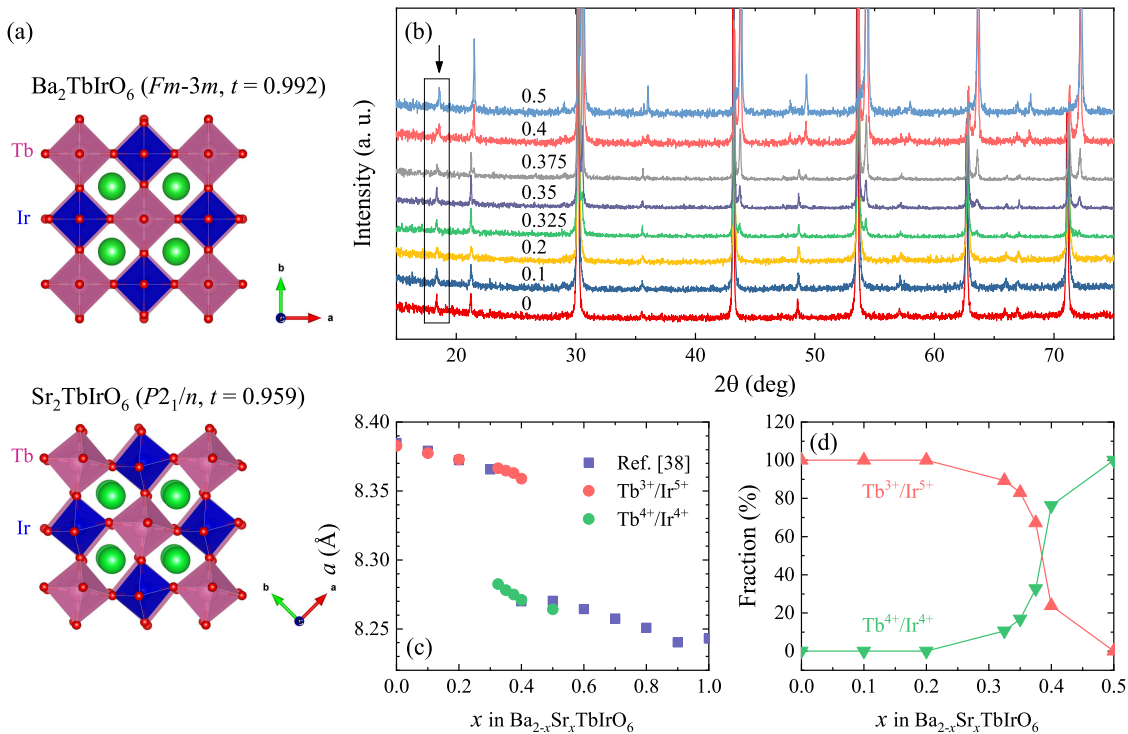}
\caption{(Color online) (a) Crystal structures of Ba$_2$TbIrO$_6$ and Sr$_2$TbIrO$_6$ projected in the $ab$ plane. The space group and tolerance factor $t$ are also given. (b) Powder x-ray diffraction patterns of Ba$_{2-x}$Sr$_x$TbIrO$_6$ collected at room temperature. The superlattice indicating the ordered arrangement of Tb and Ir cations is highlighted by the rectangle. (c) Composition dependence of the refined lattice parameters. The lattice parameters reported in Ref. [\onlinecite{zhou2005independent}] are also given for comparison. In the composition range plotted, both the Ba$_{2-x}$Sr$_x$Tb$^{3+}$Ir$^{5+}$O$_6$ and Ba$_{2-x}$Sr$_x$Tb$^{4+}$Ir$^{4+}$O$_6$ phases are cubic with a space group \textit{Fm}-3\textit{m}.(d) Evolution with $x$ of the fraction of Ba$_{2-x}$Sr$_x$Tb$^{3+}$Ir$^{5+}$O$_6$ and Ba$_{2-x}$Sr$_x$Tb$^{4+}$Ir$^{4+}$O$_6$ determined from the Rietveld refinements.}
\label{XRD_RT}
\end{figure*}

\section{Experiments}

Polycrystalline Ba$_{2-x}$Sr$_x$TbIrO$_6$ samples ($x$ = 0, 0.1, 0.2, 0.325, 0.35, 0.375, 0.4, 0.5, and 2) were synthesized by the solid state reaction method. BaCO$_3$, SrCO$_3$, Tb$_4$O$_7$, and IrO$_2$ were used as starting materials. The stoichiometric and homogeneous mixture was pelletized and fired from 1000$^\circ$C to 1200$^\circ$C for up to 6 days with intervening regrindings and repelletizings.

Powder x-ray diffraction (XRD) was carried out on a PANalytical X'Pert Pro MPD powder x-ray diffractometer using Cu $K_{\alpha 1}$ radiation. Room-temperature diffractions for all compositions were performed in the range 10$^\circ \leqslant 2\theta \leqslant 90^\circ$. Variable temperature XRD for $x$ = 0.2 was collected between 52.5$^\circ$ and 55.5$^\circ$ down to 20 K using an Oxford PheniX closed cycle cryostat.

Magnetic susceptibility ($\chi$) was measured in the temperature range 2 K $\leqslant T \leqslant$ 380 K on a Magnetic Property Measurement System (Quantum Design, QD). Specific heat ($C\rm_p$) was measured by the relaxation method between 2 K and 300 K using a QD Physical Property Measurement System (PPMS). Electrical resistivity ($\rho$) was also measured between 100 K and 380 K on the PPMS.

Neutron powder diffraction was performed on the HB-2A Powder diffractometer at the High Flux Isotope Reactor, Oak Ridge National Laboratory. The sample was contained in an annular Al sample holder to reduce the absorption from Ir. Diffraction patterns were collected from 1.5 K to 60 K in a helium cryostat. A wavelength of 2.41 \AA~  was selected from a vertically focusing germanium monochromator from the Ge113 reflection.

Resonant Inelastic x-ray Scattering (RIXS) was performed on the MERIX beamline, 27-ID, at the Advanced Photon Source, Argonne National Laboratory. The finely ground powder samples were contained in a custom Al sample holder with Kapton windows for the three compositions of $x$ = 0.1, 0.2, 2.0. The incident energy was tuned to the Ir $L_3$-edge (11.215 keV) resonant edge to enhance the Ir scattering. The inelastic energy was measured with the use of a Si(844) analyzer. The energy resolution was determined to be 35 meV at full width half maximum, based on fitting the quasi-elastic line to a charge peak. The scattering plane and incident photon polarization were both horizontal, i.e. $\pi$ incident polarization, with the incident beam focused to a size of 40 $\times$ 25 microns$^2$ (H$\times$V) at the sample position. To minimize the elastic scattering, measurements were performed with 2$\theta$ at 90$^{\circ}$ in horizontal geometry. Temperature dependent measurements were collected from 5 K to 295 K using a closed cycle refrigerator.

\section{Results and discussions}

\subsection{Valence-state transition in Ba$_{2-x}$Sr$_x$TbIrO$_6$ from room-temperature XRD}

Figure \ref{XRD_RT}(b) shows the powder XRD patterns of Ba$_{2-x}$Sr$_x$TbIrO$_6$ (0 $\leqslant x \leqslant$ 0.5) collected at room temperature. All these samples were fired in air at 1200$^\circ$C for 48 hours. The existence of the superlattice reflection for all compositions at $2\theta \approx 19^\circ$ indicates an ordered arrangement of Tb and Ir ions. For $x$ = 0, the Rietveld refinement (see Fig. S1 in Supplemental Materials) confirms the cubic structure, and the lattice parameters agree well with previous reports of Ba$_{2-x}$Sr$_x$Tb$^{3+}$Ir$^{5+}$O$_6$ [38]. With increasing Sr concentration, extra reflections are observed above $x$ = 0.325. These extra reflections can also be indexed with the cubic $Fm\overline{3}m$ structure but with a smaller lattice parameter.  According to Refs. \cite{zhou2005independent,wakeshima2000crystal}, this new cubic phase is Ba$_{2-x}$Sr$_x$Tb$^{4+}$Ir$^{4+}$O$_6$. The intensity of those reflections from Tb$^{4+}$/Ir$^{4+}$ increases with more Sr substitution. Figure \ref{XRD_RT}(c) shows the evolution with Sr content of the room temperature lattice parameters. Also plotted are the data from Ref. \cite{zhou2005independent}, and the results agree well. With increasing $x$, the lattice parameter gradually decreases and shows a sudden drop around $x_c$ = 0.325, which corresponds to the valence-state transition with electron configuration from Tb$^{3+}$/Ir$^{5+}$ to Tb$^{4+}$/Ir$^{4+}$ state. From room temperature powder XRD patterns, these two cubic phases coexist in the composition range 0.325 $\leqslant x \leqslant$ 0.4. Figure \ref{XRD_RT}(d) shows how Sr content affects the fractions of the Tb$^{3+}$/Ir$^{5+}$ and Tb$^{4+}$/Ir$^{4+}$ phases in Ba$_{2-x}$Sr$_x$TbIrO$_6$. The fraction of the Tb$^{4+}$/Ir$^{4+}$ phase is dramatically enhanced in $x$ = 0.4 with only a few percentage of residual Tb$^{3+}$/Ir$^{5+}$ phase.

We noticed that the fraction of these two cubic phases depends on the sintering conditions. For example, for $x$ = 0.35, about 17\%wt of Tb$^{4+}$/Ir$^{4+}$ phase was found when the pellet was fired at 1200$^\circ$C for 48 hours. This fraction increases to 33\%wt when fired at 1250$^\circ$C for 48 hours or at 1200$^\circ$C for 144 hours (see Fig. S2 in Supporting Materials for details). This processing dependence might be related to variation of Ir deficiency and/or the oxygen content in the sample after an extended sintering.

\subsection{Temperature-induced valence-state transition in Ba$_{1.8}$Sr$_{0.2}$TbIrO$_6$}

\begin{figure}
\includegraphics[clip,width=8.5cm]{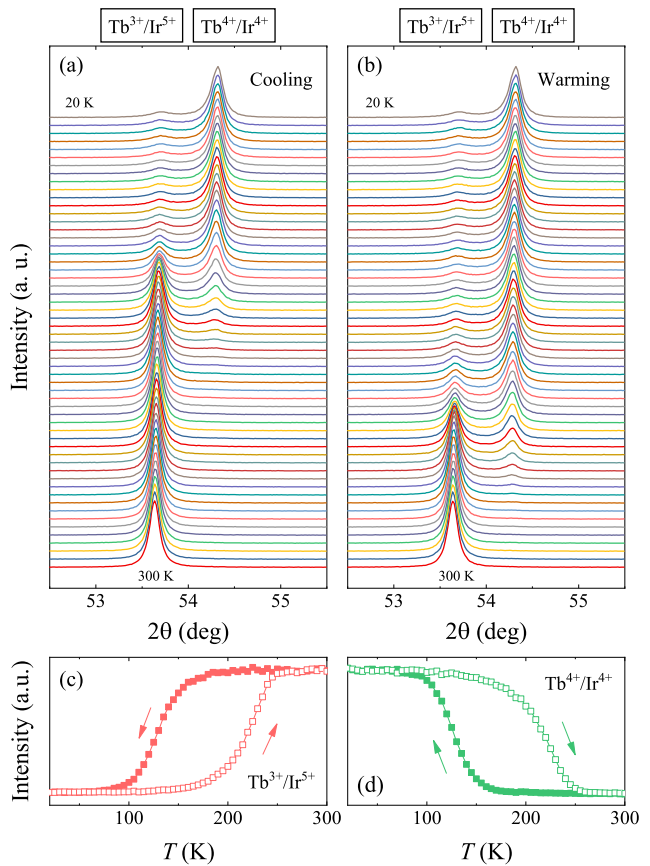}
\caption{(Color online) Evolution with temperature of the cubic (422) peak for $x$ = 0.2 monitored upon (a) cooling and (b) warming in the temperature range 20 K $\leqslant T \leqslant$ 300 K. The arrows denote the direction of sweeping temperature. (c, d) Temperature dependencies of the (422) peak intensity for Tb$^{3+}$/Ir$^{5+}$ and Tb$^{4+}$/Ir$^{4+}$ phase, respectively.}
\label{XRD_LT}
\end{figure}

The above room-temperature XRD studies show that the doping-induced valence-state transition occurs in a narrow composition range around $x_c$ = 0.325. As explained in the Introduction, one would expect a thermally driven valence-state transition near $x_c$ = 0.325 considering the temperature dependence of the tolerance factor for most perovskites. To verify this, we investigated the structure, valence state of Ir, magnetic and transport properties, magnetic structure of Ba$_{1.8}$Sr$_{0.2}$TbIrO$_6$ by measuring low-temperature x-ray and neutron powder diffraction, RIXS, magnetic susceptibility, and electrical resistivity. This composition is particularly investigated because it is single phase at room temperature as shown in Fig. \ref{XRD_RT}. Therefore, the appearance of the Tb$^{4+}$/Ir$^{4+}$ phase or the valence-state transition could be better resolved upon cooling.

\textit{Low-temperature XRD ---} The variable-temperature powder XRD patterns were recorded down to 20 K. Figures \ref{XRD_LT}(a, b) show the evolution with temperature of the (422) reflection upon cooling and warming, respectively. At 300 K, one single peak is observed at $2\theta \approx 53.6^\circ$, suggesting that the sample has a single Tb$^{3+}$/Ir$^{5+}$ phase. With decreasing temperature, a weak peak centering around $2\theta \approx 54.3^\circ$, which is the (422) reflection of the Tb$^{4+}$/Ir$^{4+}$ phase, starts to appear below $T\rm_{v, \downarrow}$ = 150 K, as seen in Fig. \ref{XRD_LT}(a). The peak intensity of Tb$^{4+}$/Ir$^{4+}$ phase increases upon further cooling, while the peak intensity of Tb$^{3+}$/Ir$^{5+}$ phase decreases. After dwelling at 20 K for about one hour, the profile of these two peaks were monitored upon warming. The peak intensity of Tb$^{4+}$/Ir$^{4+}$ phase starts to decrease around $T\rm_{v,\uparrow}$ = 180 K while that of Tb$^{3+}$/Ir$^{5+}$ phase starts to increase. Above about 250 K, only the Tb$^{3+}$/Ir$^{5+}$ phase is left. We also performed low-temperature x-ray powder diffraction measurements on \emph{x} = 0.1 and 0.5. As shown in the Supporting Materials (Fig. S3), we did not observe similar dramatic change in peak intensity.

The appearance of the Tb$^{4+}$/Ir$^{4+}$ phase at lower temperatures provides a clear evidence for the temperature-induced valence-state transition from Tb$^{3+}$/Ir$^{5+}$ to Tb$^{4+}$/Ir$^{4+}$ configuration in Ba$_{1.8}$Sr$_{0.2}$TbIrO$_6$. The different transition temperatures $T\rm_{v,\uparrow} \approx$ 180 K and $T\rm_{v,\downarrow} \approx$ 150 K indicate a first-order nature of this transition. The transition is not complete at 20 K, leading to a mixed-phase state due to the polycrystalline nature of our samples.

\begin{figure}
\includegraphics[clip,width=7.5cm]{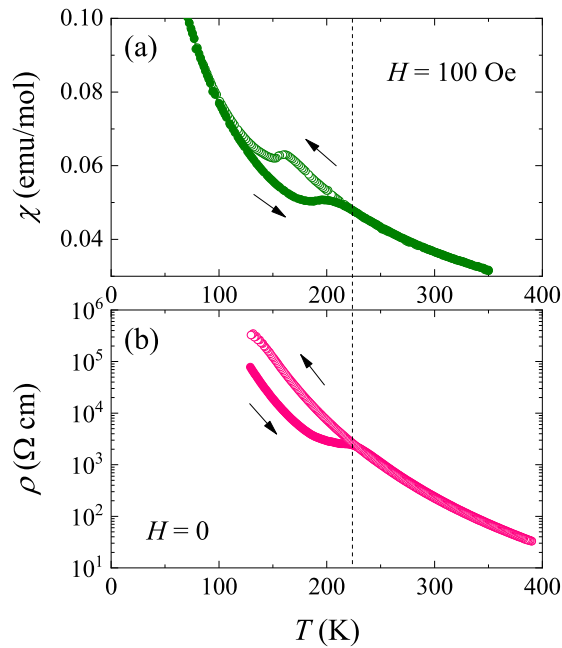}
\caption{(Color online) Temperature dependence of (a) magnetic susceptibility and (b) electrical resistivity for $x$ = 0.2 measured upon warming and cooling as marked by arrows.}
\label{MT-1}
\end{figure}

\textit{Magnetic and transport response to the valence-state transition ---} The magnetic susceptibility, $\chi(T)$, as a function of temperature measured in an applied field of 100 Oe is shown in Fig. \ref{MT-1}(a).  Upon cooling, $\chi(T)$ shows a step-like change around 170 K. While upon warming, the change occurs around 200 K. The obvious hysteresis in $\chi(T)$ is consistent with that observed in low temperature XRD measurements and is a clear evidence for the first-order nature of this valence-state transition. A Curie-Weiss fitting of the magnetic susceptibility above 200 K gives an effective moment of 9.70 $\mu\rm_B$ and a Weiss constant of -21 K. The effective moment is as expected for Tb$^{3+}$ and similar to that of $x$ = 0.1 presented later.

Figure \ref{MT-1}(b) shows the temperature dependence of the electrical resistivity, $\rho(T)$. A resistive behaviour is observed below 400 K. Upon cooling, $\rho(T)$ increases from about 20 $\Omega$cm at 400 K to 3 $\times$ 10$^5$ $\Omega$cm at 130 K, below which the sample is too resistive for any reliable data from our machine. When measured on warming, $\rho(T)$ exhibits a kink around 220 K, where the hysteresis in $\chi(T)$ occurs. We also measured $\rho(T)$ for $x$ = 0.1 and 0.5 where no valence-state transition is expected (data not shown). We did not see any anomaly as observed in Fig. \ref{MT-1}(b).

\begin{figure}
\includegraphics[clip,width=8.5cm]{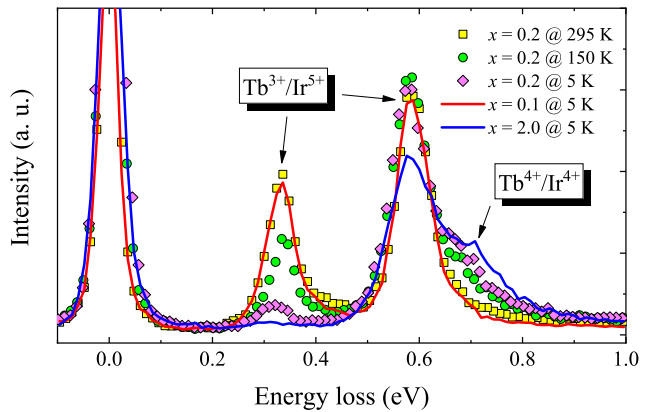}
\caption{(Color online) RIXS spectra as a function of energy loss measured for $x$ = 0.2. Data collected on $x$ = 0.1 and $x$ = 2 are used as a reference for Tb$^{3+}$/Ir$^{5+}$ and Tb$^{4+}$/Ir$^{4+}$ configurations, respectively.}
\label{XAS}
\end{figure}

\textit{Resonant inelastic x-ray scattering ---} To verify that the above anomalies in the low-temperature XRD study is indeed from the transition between Tb$^{3+}$/Ir$^{5+}$ and Tb$^{4+}$/Ir$^{4+}$ phases due to the charge transfer between Tb and Ir ions, we measured RIXS for three Ba$_{2-x}$Sr$_x$TbIrO$_6$ samples with $x$ = 0.1, 0.2, and 2. RIXS is element specific and measures directly transitions between $d-d$ orbitals; therefore, it is a powerful probe to provide direct information on the valence states of the Ir ions in this compound \cite{ament2011resonant}. At 295 K, as shown in Fig. \ref{XAS}, the spectrum for $x$ = 0.2 exhibits two peaks around 0.3 and 0.6 eV. At lower temperatures, two features are observed: (i) the intensity of the high-energy peak is nearly temperature independent while the low-energy peak is significantly suppressed; (ii) a shoulder around 0.7 eV gradually arises. To explain the evolution of the peak intensity with temperature, the RIXS spectra of $x$ = 0.1 and 2 are also plotted as a reference. At 295 K, the spectrum of $x$ = 0.2 matches well with the $x$ = 0.1 curve, indicating a pure Tb$^{3+}$/Ir$^{5+}$ phase. At lower temperatures, the emergence of the shoulder feature indicates the presence of the Tb$^{4+}$/Ir$^{4+}$ phase, which is therefore a very direct evidence for the valence-state transition between Tb$^{3+}$/Ir$^{5+}$ and Tb$^{4+}$/Ir$^{4+}$ states. The weak but visible low-energy peak for $x$ = 0.2 at 5 K supports the phase coexistence as deduced from the variable-temperature XRD results.

\begin{figure}
\includegraphics[clip,width=7.5cm]{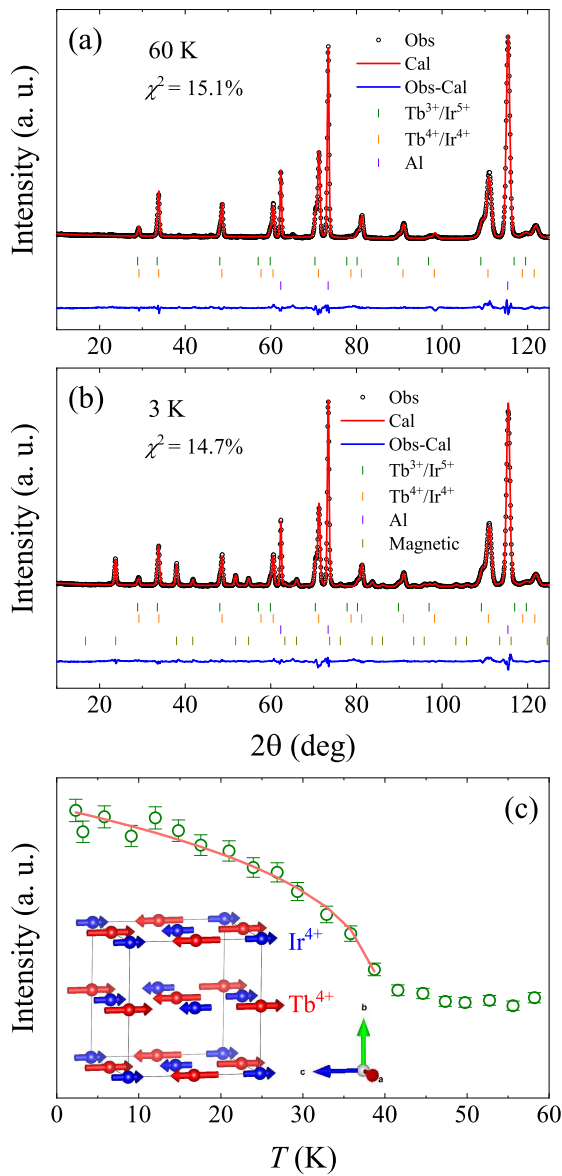}
\caption{(Color online) Neutron powder diffraction performed on $x$ = 0.2 at (a) 60 K and (b) 3 K. In panel (a), the vertical-tick marks from top to bottom correspond to Tb$^{3+}$/Ir$^{5+}$ phase (green), Tb$^{4+}$/Ir$^{4+}$ phase (orange), and Al can (blue). The additional set of vertical tick marks in panel (b) at the bottom are the magnetic reflections. (c) Temperature dependence of the peak intensity around $2\theta$ = 23.8$^\circ$. The solid line is a fit as described in the text. The inset is the arrangement of Tb$^{4+}$ spins (red) and Ir$^{4+}$ spins (blue).}
\label{Neutron}
\end{figure}

\begin{table*}[!ht]
\caption{Structural parameters of Ba$_{2-x}$Sr$_x$TbIrO$_6$ ($x$ = 0.2) at different temperatures obtained from Rietveld refinement of neutron powder diffraction patterns. The coexistence of the Tb$^{3+}$/Ir$^{5+}$ and Tb$^{4+}$/Ir$^{4+}$ phases with a reasonable phase fraction allows us to extract the structural parameters of each phase. The space group is \textit{Fm}-3\textit{m} for both phases. The position of oxygen (\textit{x}, 0, 0) is presented. Refinement of the Ba/Sr ratio found agreement with the nominal composition. The refinement also suggests a stoichiometric amount of oxygen. The oxidization states of Tb and Ir estimated from bond valence sum (BVS) are also shown. For the BVS calculation\cite{brese1991bond}, Tb$^{4+}$, R$_0$=2.018\AA~ and B=0.395. For Ir$^{4+}$, R$_0$=1.909\AA~ and B=0.258.}
\label{table1}
\begin{ruledtabular}
\begin{tabular} {lllllll}
& \multicolumn{3}{c} {Tb$^{3+}$/Ir$^{5+}$ phase} & \multicolumn{3}{c} {Tb$^{4+}$/Ir$^{4+}$ phase}\\[0.5ex]
& 250 K & 165 K & 60 K & 250 K & 165 K & 60 K \\
\cline{2-4} \cline{5-7}\\[0.2ex]
$a$ ({\AA}) & 8.3754(2) & 8.3737(2) & 8.3715(3) & & 8.2920(2) & 8.2885(2) \\
$x_O$ & 0.2344(2) & 0.2328(4) & 0.2329(6) & & 0.2407(3) & 0.2412(3) \\
Ba/Sr-O ({\AA}) & 2.96402(10) & 2.96405(18) & 2.9632(3) & & 2.93268(10) & 2.93132(10) \\
Tb-O ({\AA}) & 2.2245(17) & 2.237(4) & 2.236(6) & & 2.150(3) & 2.145(3) \\
BVS Tb$^{4+}$ & 3.56 & 3.45 & 3.46 & &4.30 & 4.35 \\
Ir-O ({\AA}) & 1.9632(17) & 1.949(4) & 1.950(6) & & 1.996(3) & 1.999(3) \\
BVS Ir$^{4+}$ & 4.86 & 5.14 & 5.12 & &4.28 & 4.23 \\
$\angle$Tb-O-Ir (deg) & 180 & 180 & 180 & & 180 & 180 \\
Fraction & 100\% & 34.7(4)\% & 22.8(3)\% & & 65.3(6)\% & 77.2\% \\
\end{tabular}
\end{ruledtabular}
\end{table*}

\textit{Nuclear and magnetic structure from neutron powder diffraction}--- Figure \ref{Neutron}(a) shows the Rietveld refinement of the neutron powder diffraction pattern for $x$ = 0.2 collected at 60 K, which is consistent with the XRD result including both Tb$^{3+}$/Ir$^{5+}$ and Tb$^{4+}$/Ir$^{4+}$ phases below $T\rm_v$. The respective lattice parameter is $a$ = 8.3423(5) {\AA} for the Tb$^{3+}$/Ir$^{5+}$ phase and $a$ = 8.2680(2) {\AA} for the Tb$^{4+}$/Ir$^{4+}$ phase. Some extra strong reflections are observed at 3 K in Fig. \ref{Neutron}(b). The $d$-spacing proves that this set of magnetic peaks come from the Tb$^{4+}$/Ir$^{4+}$ phase. Figure \ref{Neutron}(c) shows the temperature dependence of the peak intensity at $2\theta = 23.8^\circ$, which confirms the long-range magnetic order at 40 K. The positions of the magnetic reflections are compatible with a propagation vector \textbf{k} = (0, 0, 1). A representational analysis approach was utilized with the aid of the SARAh program \cite{wills2000new}. For a propagation vector \textbf{k} = (0, 0, 1) and magnetic Tb ions at (0.5, 0.5, 0.5) and Ir ions at (0, 0, 0) in the $Fm\overline{3}m$ space group, there are two symmetry allowed irreducible representations: $\Gamma_3$ and $\Gamma_9$ in Kovalevs scheme. $\Gamma_9$ confines the spins on both Ir and Tb ions to the $c$ axis and was found to provide the best fit to the data. Any determination of canting away from the $c$ axis is beyond the limits of the data. As illustrated in the inset to Fig. \ref{Neutron}(c), both the magnetic moments of the Tb$^{4+}$ and Ir$^{4+}$ spins are aligned along the $c$ direction. The Tb$^{4+}$ and Ir$^{4+}$ sublattices couple ferromagnetically in the $ab$ plane, which are further arranged antiferromagnetically along the $c$ direction. The deduced magnetic moment is about 5.98(4) $\mu\rm_B$ for Tb$^{4+}$ ion and 0.5(1) $\mu\rm_B$ for Ir$^{4+}$ ion. It should be noted that the low signal from Ir makes accurate isolation of the signal challenging and these values are best fits only.

The coexistence of the Tb$^{3+}$/Ir$^{5+}$ and Tb$^{4+}$/Ir$^{4+}$ phases with a reasonable phase fraction allows us to obtain the structural parameters of each phase from Rietveld refinement of the neutron powder diffraction patterns collected at different temperatures. As shown in Table I, the Tb$^{4+}$/Ir$^{4+}$ phase has a smaller lattice parameter than the Tb$^{3+}$/Ir$^{5+}$ phase. This is consistent with that determined by x-ray powder diffraction. The  Tb$^{4+}$/Ir$^{4+}$ phase has an elongated Ir-O bond but shortened Ba-O and Tb-O bonds compared to the  Tb$^{3+}$/Ir$^{5+}$ phase. Also shown in Table I are the oxidization states of Tb and Ir estimated from bond valence sum, which in general agree with the expected oxidization states of Tb and Ir ions in each phase.

\begin{figure}
\includegraphics[clip,width=7.25cm]{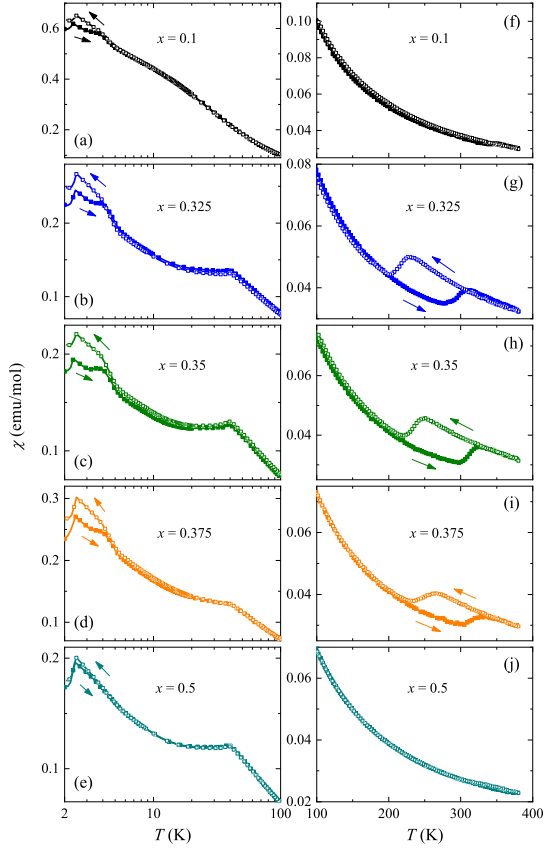}
\caption{(Color online) Temperature dependence of magnetic susceptibility for Ba$_{2-x}$Sr$_x$TbIrO$_6$ series measured in $H$ = 100 Oe. $x$ = 0.5 sample was measured in 500 Oe. Panels (a-f) highlight the features at low temperatures. The magnetic susceptibility of \textit{x} = 0.20 is similar to that of \textit{x} = 0.325 shown in panel (b). Panels (g-l) highlight the valence-state transition at high temperatures. Solid (open) symbols are measured in ZFC (FC) process. The drop of the magnetic susceptibility near 3 K comes from a small amount of Tb$_2$O$_3$ impurity.}
\label{MT-2}
\end{figure}

\subsection{Evolution with $x$ of the valence-state transition and low-temperature magnetic order of Ba$_{2-x}$Sr$_x$TbIrO$_6$}

In order to understand how the valence-state transition evolves with the chemical pressure, we measured $\chi(T)$ curves of six different Ba$_{2-x}$Sr$_x$TbIrO$_6$ samples with 0.1 $\leqslant x \leqslant$ 0.5. Figures \ref{MT-2}(a-e) show the $\chi(T)$ data below 100 K. For $x$ = 0.1, $\chi(T)$ increases upon cooling and exhibits a slope change around 5 K, below which $\chi(T)$ measured in field-cooling (FC) mode starts to diverge from that collected in zero-field-cooling (ZFC). For $x$ = 0.5, the slope change is absent but a kink at 40 K can be well resolved. For other compositions, both the slope change around 5 K and the kink around 40 K are present. A sudden drop of $\chi(T)$ upon cooling across $\approx$ 2.5 K is observed in all compositions and comes from a small amount of Tb$_2$O$_3$ impurity \cite{veber2015flux} that can barely be observed by room temperature x-ray powder diffraction. At high temperatures, see Figs. \ref{MT-2}(f-j), a paramagnetic behavior is observed for $x$ = 0.1 and 0.5 up to 380 K, while a clear first-order transition is observed for 0.2 $\leqslant x \leqslant$ 0.375 in the temperature range 180 K $\leqslant T \leqslant$ 320 K. The valence-state transition shifts to higher temperatures with increasing $x$. A Curie-Weiss fitting of the high-temperature magnetic susceptibility for $x$ = 0.1 gives an effective moment of 9.74 $\mu\rm_B$ and a Weiss constant of -16 K. This effective moment is as expected for Tb$^{3+}$ ions and confirms the Tb$^{3+}$/Ir$^{5+}$ state for $x$ = 0.1. The Curie-Weiss fitting of $x$ = 0.5 gives an effective moment of 8.47 $\mu\rm_B$ and a Weiss constant of -30 K. This effective moment is slightly larger than the expected value of 8.13 $\mu\rm_B$ for Tb$^{4+}$/Ir$^{4+}$ configuration.

The above complex temperature dependence of $\chi(T)$ shows a close relation between the high-temperature valence-state transition with the low-temperature magnetic order. For $x$ = 0.1 with only Tb$^{3+}$/Ir$^{5+}$ phase, no long-range magnetic order is observed but a spin-glass state is developed below $\approx$ 5 K. For $x$ = 0.5 with only Tb$^{4+}$/Ir$^{4+}$ phase, the kink feature around 40 K is a signature for the long-range AFM order. The absence of the splitting between FC and ZFC $\chi(T)$ indicates a collinear AFM spin arrangement as confirmed for the Tb$^{4+}$/Ir$^{4+}$ phase in $x$ = 0.2 by the neutron diffraction measurements. For other compositions with a valence-state transition, the incomplete transition results in the coexistence of Tb$^{3+}$/Ir$^{5+}$ and Tb$^{4+}$/Ir$^{4+}$ phases. Therefore, both the slope change around 5 K and the kink around 40 K  are observed in $\chi(T)$.

\begin{figure}
\includegraphics[clip,width=8.5cm]{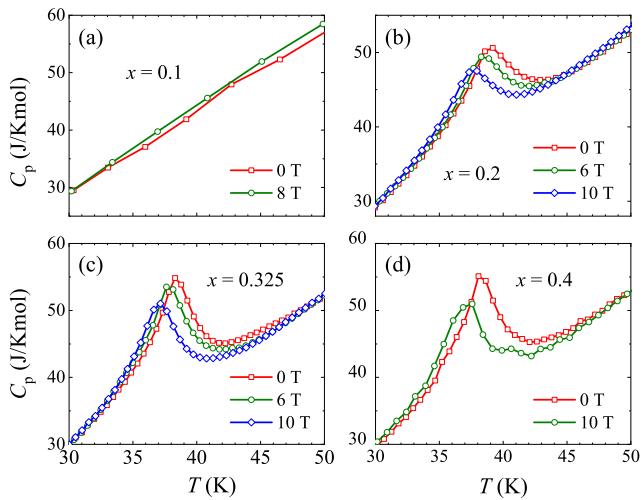}
\caption{(Color online) Specific heat of selected compositions of Ba$_{2-x}$Sr$_x$TbIrO$_6$.}
\label{Cp}
\end{figure}

The magnetic order around 40 K can be well resolved in the temperature dependence of specific heat as shown in Fig. \ref{Cp}. For $x$ = 0.1, no anomaly is observed at either around 5 K or 40 K. The former is consistent with the spin-glass origin of the splitting of FC and ZFC curves; the latter confirms the absence of a long-range magnetic order. For other compositions, a $\lambda$-type anomaly due to the AFM order of the Tb$^{4+}$/Ir$^{4+}$ phase shows up near $T\rm_N$ = 40 K. Application of a magnetic field can suppress the magnetic order and reduce $T\rm_N$.

\section{Summary}

In summary, we study the temperature-induced valence-state transition in a narrow composition range 0.2 $\leqslant x \leqslant$ 0.375 in Ba$_{2-x}$Sr$_x$TbIrO$_6$. Upon cooling, one electron is transferred from Tb$^{3+}$ to Ir$^{5+}$ in the Ba$_{2-x}$Sr$_x$Tb$^{3+}$Ir$^{5+}$O$_6$ phase leading to the formation of the Ba$_{2-x}$Sr$_x$Tb$^{4+}$Ir$^{4+}$O$_6$ phase. No long-range magnetic order was observed in Ba$_{2-x}$Sr$_x$Tb$^{3+}$Ir$^{5+}$O$_6$ where the Ir$^{5+}$ has an electronic configuration of 5$d^4$ ($J\rm_{eff}$ = 0). The observation of the long-range magnetic order with $T\rm_N$ = 40 K in Ba$_{2-x}$Sr$_x$Tb$^{4+}$Ir$^{4+}$O$_6$, at which both Tb$^{4+}$ and Ir$^{4+}$ ions order simultaneously, suggests an essential role of Ir$^{4+}$ ($J\rm_{eff}$ = 1/2) in mediating the magnetic interactions of Tb ions.

We noticed that the valence-state transition is not complete even at 2 K in our polycrystalline samples. Detailed studies on single crystals are desired to further understand the mechanism and effects of the valence-state transition. The first-order valence-state transition is sensitive to the chemical pressure induced by the size difference between Ba$^{2+}$ and Sr$^{2+}$ cations. It is thus reasonable to expect that strain field can be effective in controlling the valence-state transition and thus the magnetism. This kind of effect can be better tested in thin films on different substrates or in-situ strain fields. Our results reported in this work show that the valence-state transition has a dramatic effect on the structure, magnetic and transport properties. Appropriate chemical doping that might induce electrical conductivity also deserves some efforts for novel phenomena accompanying with the valence-state transition.

\section{Acknowledgement}

JQY would like to thank Nandini Trivedi, Patrick Woodward, and Jianshi Zhou for helpful discussions. Work at ORNL was supported by the US Department of Energy, Office of Science, Basic Energy Sciences, Materials Sciences and Engineering Division. Synthesis of powder samples for neutron diffraction measurement and part of the manuscript preparation by ZZ were supported by the National Natural Science Foundation of China (Grants No. U1832166 and 52072368). This research used resources at the High Flux Isotope Reactor, a DOE office of Science User Facility operated by the Oak Ridge National Laboratory. This research used resources of the Advanced Photon Source, a U.S. Department of Energy (DOE) Office of Science User Facility operated for the DOE Office of Science by Argonne National Laboratory under Contract No. DE-AC02-06CH11357.

\section{references}


\begin{thebibliography}{45}%
\makeatletter
\providecommand \@ifxundefined [1]{%
 \@ifx{#1\undefined}
}%
\providecommand \@ifnum [1]{%
 \ifnum #1\expandafter \@firstoftwo
 \else \expandafter \@secondoftwo
 \fi
}%
\providecommand \@ifx [1]{%
 \ifx #1\expandafter \@firstoftwo
 \else \expandafter \@secondoftwo
 \fi
}%
\providecommand \natexlab [1]{#1}%
\providecommand \enquote  [1]{``#1''}%
\providecommand \bibnamefont  [1]{#1}%
\providecommand \bibfnamefont [1]{#1}%
\providecommand \citenamefont [1]{#1}%
\providecommand \href@noop [0]{\@secondoftwo}%
\providecommand \href [0]{\begingroup \@sanitize@url \@href}%
\providecommand \@href[1]{\@@startlink{#1}\@@href}%
\providecommand \@@href[1]{\endgroup#1\@@endlink}%
\providecommand \@sanitize@url [0]{\catcode `\\12\catcode `\$12\catcode
  `\&12\catcode `\#12\catcode `\^12\catcode `\_12\catcode `\%12\relax}%
\providecommand \@@startlink[1]{}%
\providecommand \@@endlink[0]{}%
\providecommand \url  [0]{\begingroup\@sanitize@url \@url }%
\providecommand \@url [1]{\endgroup\@href {#1}{\urlprefix }}%
\providecommand \urlprefix  [0]{URL }%
\providecommand \Eprint [0]{\href }%
\providecommand \doibase [0]{http://dx.doi.org/}%
\providecommand \selectlanguage [0]{\@gobble}%
\providecommand \bibinfo  [0]{\@secondoftwo}%
\providecommand \bibfield  [0]{\@secondoftwo}%
\providecommand \translation [1]{[#1]}%
\providecommand \BibitemOpen [0]{}%
\providecommand \bibitemStop [0]{}%
\providecommand \bibitemNoStop [0]{.\EOS\space}%
\providecommand \EOS [0]{\spacefactor3000\relax}%
\providecommand \BibitemShut  [1]{\csname bibitem#1\endcsname}%
\let\auto@bib@innerbib\@empty
\bibitem [{\citenamefont {Witczak-Krempa}\ \emph {et~al.}(2014)\citenamefont
  {Witczak-Krempa}, \citenamefont {Chen}, \citenamefont {Kim},\ and\
  \citenamefont {Balents}}]{witczak2014correlated}%
  \BibitemOpen
  \bibfield  {author} {\bibinfo {author} {\bibfnamefont {William}\ \bibnamefont
  {Witczak-Krempa}}, \bibinfo {author} {\bibfnamefont {Gang}\ \bibnamefont
  {Chen}}, \bibinfo {author} {\bibfnamefont {Yong~Baek}\ \bibnamefont {Kim}}, \
  and\ \bibinfo {author} {\bibfnamefont {Leon}\ \bibnamefont {Balents}},\
  }\bibfield  {title} {\enquote {\bibinfo {title} {Correlated quantum phenomena
  in the strong spin-orbit regime},}\ }\href@noop {} {\bibfield  {journal}
  {\bibinfo  {journal} {Annu. Rev. Condens. Matter Phys.}\ }\textbf {\bibinfo
  {volume} {5}},\ \bibinfo {pages} {57--82} (\bibinfo {year}
  {2014})}\BibitemShut {NoStop}%
\bibitem [{\citenamefont {Rau}\ \emph {et~al.}(2016)\citenamefont {Rau},
  \citenamefont {Lee},\ and\ \citenamefont {Kee}}]{rau2016spin}%
  \BibitemOpen
  \bibfield  {author} {\bibinfo {author} {\bibfnamefont {Jeffrey~G}\
  \bibnamefont {Rau}}, \bibinfo {author} {\bibfnamefont {Eric Kin-Ho}\
  \bibnamefont {Lee}}, \ and\ \bibinfo {author} {\bibfnamefont {Hae-Young}\
  \bibnamefont {Kee}},\ }\bibfield  {title} {\enquote {\bibinfo {title}
  {Spin-orbit physics giving rise to novel phases in correlated systems:
  Iridates and related materials},}\ }\href@noop {} {\bibfield  {journal}
  {\bibinfo  {journal} {Annual Review of Condensed Matter Physics}\ }\textbf
  {\bibinfo {volume} {7}},\ \bibinfo {pages} {195--221} (\bibinfo {year}
  {2016})}\BibitemShut {NoStop}%
\bibitem [{\citenamefont {Schaffer}\ \emph {et~al.}(2016)\citenamefont
  {Schaffer}, \citenamefont {Lee}, \citenamefont {Yang},\ and\ \citenamefont
  {Kim}}]{schaffer2016recent}%
  \BibitemOpen
  \bibfield  {author} {\bibinfo {author} {\bibfnamefont {Robert}\ \bibnamefont
  {Schaffer}}, \bibinfo {author} {\bibfnamefont {Eric Kin-Ho}\ \bibnamefont
  {Lee}}, \bibinfo {author} {\bibfnamefont {Bohm-Jung}\ \bibnamefont {Yang}}, \
  and\ \bibinfo {author} {\bibfnamefont {Yong~Baek}\ \bibnamefont {Kim}},\
  }\bibfield  {title} {\enquote {\bibinfo {title} {Recent progress on
  correlated electron systems with strong spin--orbit coupling},}\ }\href@noop
  {} {\bibfield  {journal} {\bibinfo  {journal} {Reports on Progress in
  Physics}\ }\textbf {\bibinfo {volume} {79}},\ \bibinfo {pages} {094504}
  (\bibinfo {year} {2016})}\BibitemShut {NoStop}%
\bibitem [{\citenamefont {Kim}\ \emph {et~al.}(2008)\citenamefont {Kim},
  \citenamefont {Jin}, \citenamefont {Moon}, \citenamefont {Kim}, \citenamefont
  {Park}, \citenamefont {Leem}, \citenamefont {Yu}, \citenamefont {Noh},
  \citenamefont {Kim}, \citenamefont {Oh} \emph {et~al.}}]{kim2008novel}%
  \BibitemOpen
  \bibfield  {author} {\bibinfo {author} {\bibfnamefont {BJ}~\bibnamefont
  {Kim}}, \bibinfo {author} {\bibfnamefont {Hosub}\ \bibnamefont {Jin}},
  \bibinfo {author} {\bibfnamefont {SJ}~\bibnamefont {Moon}}, \bibinfo {author}
  {\bibfnamefont {J-Y}\ \bibnamefont {Kim}}, \bibinfo {author} {\bibfnamefont
  {B-G}\ \bibnamefont {Park}}, \bibinfo {author} {\bibfnamefont
  {CS}~\bibnamefont {Leem}}, \bibinfo {author} {\bibfnamefont {Jaejun}\
  \bibnamefont {Yu}}, \bibinfo {author} {\bibfnamefont {TW}~\bibnamefont
  {Noh}}, \bibinfo {author} {\bibfnamefont {C}~\bibnamefont {Kim}}, \bibinfo
  {author} {\bibfnamefont {S-J}\ \bibnamefont {Oh}},  \emph {et~al.},\
  }\bibfield  {title} {\enquote {\bibinfo {title} {Novel j eff= 1/2 mott state
  induced by relativistic spin-orbit coupling in sr 2 iro 4},}\ }\href@noop {}
  {\bibfield  {journal} {\bibinfo  {journal} {Physical review letters}\
  }\textbf {\bibinfo {volume} {101}},\ \bibinfo {pages} {076402} (\bibinfo
  {year} {2008})}\BibitemShut {NoStop}%
\bibitem [{\citenamefont {Takagi}\ \emph {et~al.}(2019)\citenamefont {Takagi},
  \citenamefont {Takayama}, \citenamefont {Jackeli}, \citenamefont
  {Khaliullin},\ and\ \citenamefont {Nagler}}]{takagi2019concept}%
  \BibitemOpen
  \bibfield  {author} {\bibinfo {author} {\bibfnamefont {Hidenori}\
  \bibnamefont {Takagi}}, \bibinfo {author} {\bibfnamefont {Tomohiro}\
  \bibnamefont {Takayama}}, \bibinfo {author} {\bibfnamefont {George}\
  \bibnamefont {Jackeli}}, \bibinfo {author} {\bibfnamefont {Giniyat}\
  \bibnamefont {Khaliullin}}, \ and\ \bibinfo {author} {\bibfnamefont
  {Stephen~E}\ \bibnamefont {Nagler}},\ }\bibfield  {title} {\enquote {\bibinfo
  {title} {Concept and realization of kitaev quantum spin liquids},}\
  }\href@noop {} {\bibfield  {journal} {\bibinfo  {journal} {Nature Reviews
  Physics}\ }\textbf {\bibinfo {volume} {1}},\ \bibinfo {pages} {264--280}
  (\bibinfo {year} {2019})}\BibitemShut {NoStop}%
\bibitem [{\citenamefont {Schnelle}\ \emph {et~al.}(2021)\citenamefont
  {Schnelle}, \citenamefont {Prasad}, \citenamefont {Felser}, \citenamefont
  {Jansen}, \citenamefont {Komleva}, \citenamefont {Streltsov}, \citenamefont
  {Mazin}, \citenamefont {Khalyavin}, \citenamefont {Manuel}, \citenamefont
  {Pal} \emph {et~al.}}]{schnelle2021magnetic}%
  \BibitemOpen
  \bibfield  {author} {\bibinfo {author} {\bibfnamefont {Walter}\ \bibnamefont
  {Schnelle}}, \bibinfo {author} {\bibfnamefont {Beluvalli~E}\ \bibnamefont
  {Prasad}}, \bibinfo {author} {\bibfnamefont {Claudia}\ \bibnamefont
  {Felser}}, \bibinfo {author} {\bibfnamefont {Martin}\ \bibnamefont {Jansen}},
  \bibinfo {author} {\bibfnamefont {Evgenia~V}\ \bibnamefont {Komleva}},
  \bibinfo {author} {\bibfnamefont {Sergey~V}\ \bibnamefont {Streltsov}},
  \bibinfo {author} {\bibfnamefont {Igor~I}\ \bibnamefont {Mazin}}, \bibinfo
  {author} {\bibfnamefont {Dmitry}\ \bibnamefont {Khalyavin}}, \bibinfo
  {author} {\bibfnamefont {Pascal}\ \bibnamefont {Manuel}}, \bibinfo {author}
  {\bibfnamefont {Sukanya}\ \bibnamefont {Pal}},  \emph {et~al.},\ }\bibfield
  {title} {\enquote {\bibinfo {title} {Magnetic and electronic ordering
  phenomena in the ru 2 o 6-layer honeycomb lattice compound agruo 3},}\
  }\href@noop {} {\bibfield  {journal} {\bibinfo  {journal} {Physical Review
  B}\ }\textbf {\bibinfo {volume} {103}},\ \bibinfo {pages} {214413} (\bibinfo
  {year} {2021})}\BibitemShut {NoStop}%
\bibitem [{\citenamefont {Tian}\ \emph {et~al.}(2015)\citenamefont {Tian},
  \citenamefont {Svoboda}, \citenamefont {Ochi}, \citenamefont {Matsuda},
  \citenamefont {Cao}, \citenamefont {Cheng}, \citenamefont {Sales},
  \citenamefont {Mandrus}, \citenamefont {Arita}, \citenamefont {Trivedi} \emph
  {et~al.}}]{tian2015high}%
  \BibitemOpen
  \bibfield  {author} {\bibinfo {author} {\bibfnamefont {Wei}\ \bibnamefont
  {Tian}}, \bibinfo {author} {\bibfnamefont {Chris}\ \bibnamefont {Svoboda}},
  \bibinfo {author} {\bibfnamefont {M}~\bibnamefont {Ochi}}, \bibinfo {author}
  {\bibfnamefont {M}~\bibnamefont {Matsuda}}, \bibinfo {author} {\bibfnamefont
  {HB}~\bibnamefont {Cao}}, \bibinfo {author} {\bibfnamefont {J-G}\
  \bibnamefont {Cheng}}, \bibinfo {author} {\bibfnamefont {BC}~\bibnamefont
  {Sales}}, \bibinfo {author} {\bibfnamefont {DG}~\bibnamefont {Mandrus}},
  \bibinfo {author} {\bibfnamefont {R}~\bibnamefont {Arita}}, \bibinfo {author}
  {\bibfnamefont {Nandini}\ \bibnamefont {Trivedi}},  \emph {et~al.},\
  }\bibfield  {title} {\enquote {\bibinfo {title} {High antiferromagnetic
  transition temperature of the honeycomb compound srru 2 o 6},}\ }\href@noop
  {} {\bibfield  {journal} {\bibinfo  {journal} {Physical Review B}\ }\textbf
  {\bibinfo {volume} {92}},\ \bibinfo {pages} {100404} (\bibinfo {year}
  {2015})}\BibitemShut {NoStop}%
\bibitem [{\citenamefont {Hiley}\ \emph {et~al.}(2015)\citenamefont {Hiley},
  \citenamefont {Scanlon}, \citenamefont {Sokol}, \citenamefont {Woodley},
  \citenamefont {Ganose}, \citenamefont {Sangiao}, \citenamefont {De~Teresa},
  \citenamefont {Manuel}, \citenamefont {Khalyavin}, \citenamefont {Walker}
  \emph {et~al.}}]{hiley2015antiferromagnetism}%
  \BibitemOpen
  \bibfield  {author} {\bibinfo {author} {\bibfnamefont {CI}~\bibnamefont
  {Hiley}}, \bibinfo {author} {\bibfnamefont {DO}~\bibnamefont {Scanlon}},
  \bibinfo {author} {\bibfnamefont {AA}~\bibnamefont {Sokol}}, \bibinfo
  {author} {\bibfnamefont {SM}~\bibnamefont {Woodley}}, \bibinfo {author}
  {\bibfnamefont {AM}~\bibnamefont {Ganose}}, \bibinfo {author} {\bibfnamefont
  {S}~\bibnamefont {Sangiao}}, \bibinfo {author} {\bibfnamefont
  {JM}~\bibnamefont {De~Teresa}}, \bibinfo {author} {\bibfnamefont
  {P}~\bibnamefont {Manuel}}, \bibinfo {author} {\bibfnamefont
  {DD}~\bibnamefont {Khalyavin}}, \bibinfo {author} {\bibfnamefont
  {M}~\bibnamefont {Walker}},  \emph {et~al.},\ }\bibfield  {title} {\enquote
  {\bibinfo {title} {Antiferromagnetism at t> 500 k in the layered hexagonal
  ruthenate srr u 2 o 6},}\ }\href@noop {} {\bibfield  {journal} {\bibinfo
  {journal} {Physical Review B}\ }\textbf {\bibinfo {volume} {92}},\ \bibinfo
  {pages} {104413} (\bibinfo {year} {2015})}\BibitemShut {NoStop}%
\bibitem [{\citenamefont {Shi}\ \emph {et~al.}(2009)\citenamefont {Shi},
  \citenamefont {Guo}, \citenamefont {Yu}, \citenamefont {Arai}, \citenamefont
  {Belik}, \citenamefont {Sato}, \citenamefont {Yamaura}, \citenamefont
  {Takayama-Muromachi}, \citenamefont {Tian}, \citenamefont {Yang} \emph
  {et~al.}}]{shi2009continuous}%
  \BibitemOpen
  \bibfield  {author} {\bibinfo {author} {\bibfnamefont {YG}~\bibnamefont
  {Shi}}, \bibinfo {author} {\bibfnamefont {YF}~\bibnamefont {Guo}}, \bibinfo
  {author} {\bibfnamefont {S}~\bibnamefont {Yu}}, \bibinfo {author}
  {\bibfnamefont {M}~\bibnamefont {Arai}}, \bibinfo {author} {\bibfnamefont
  {AA}~\bibnamefont {Belik}}, \bibinfo {author} {\bibfnamefont {A}~\bibnamefont
  {Sato}}, \bibinfo {author} {\bibfnamefont {K}~\bibnamefont {Yamaura}},
  \bibinfo {author} {\bibfnamefont {E}~\bibnamefont {Takayama-Muromachi}},
  \bibinfo {author} {\bibfnamefont {HF}~\bibnamefont {Tian}}, \bibinfo {author}
  {\bibfnamefont {HX}~\bibnamefont {Yang}},  \emph {et~al.},\ }\bibfield
  {title} {\enquote {\bibinfo {title} {Continuous metal-insulator transition of
  the antiferromagnetic perovskite naoso 3},}\ }\href@noop {} {\bibfield
  {journal} {\bibinfo  {journal} {Physical Review B}\ }\textbf {\bibinfo
  {volume} {80}},\ \bibinfo {pages} {161104} (\bibinfo {year}
  {2009})}\BibitemShut {NoStop}%
\bibitem [{\citenamefont {Rodriguez}\ \emph {et~al.}(2011)\citenamefont
  {Rodriguez}, \citenamefont {Poineau}, \citenamefont {Llobet}, \citenamefont
  {Kennedy}, \citenamefont {Avdeev}, \citenamefont {Thorogood}, \citenamefont
  {Carter}, \citenamefont {Seshadri}, \citenamefont {Singh},\ and\
  \citenamefont {Cheetham}}]{rodriguez2011high}%
  \BibitemOpen
  \bibfield  {author} {\bibinfo {author} {\bibfnamefont {Efrain~E}\
  \bibnamefont {Rodriguez}}, \bibinfo {author} {\bibfnamefont
  {Fr{\'e}d{\'e}ric}\ \bibnamefont {Poineau}}, \bibinfo {author} {\bibfnamefont
  {Anna}\ \bibnamefont {Llobet}}, \bibinfo {author} {\bibfnamefont {Brendan~J}\
  \bibnamefont {Kennedy}}, \bibinfo {author} {\bibfnamefont {Maxim}\
  \bibnamefont {Avdeev}}, \bibinfo {author} {\bibfnamefont {Gordon~J}\
  \bibnamefont {Thorogood}}, \bibinfo {author} {\bibfnamefont {Melody~L}\
  \bibnamefont {Carter}}, \bibinfo {author} {\bibfnamefont {Ram}\ \bibnamefont
  {Seshadri}}, \bibinfo {author} {\bibfnamefont {David~J}\ \bibnamefont
  {Singh}}, \ and\ \bibinfo {author} {\bibfnamefont {Anthony~K}\ \bibnamefont
  {Cheetham}},\ }\bibfield  {title} {\enquote {\bibinfo {title} {High
  temperature magnetic ordering in the 4 d perovskite srtco 3},}\ }\href@noop
  {} {\bibfield  {journal} {\bibinfo  {journal} {Physical review letters}\
  }\textbf {\bibinfo {volume} {106}},\ \bibinfo {pages} {067201} (\bibinfo
  {year} {2011})}\BibitemShut {NoStop}%
\bibitem [{\citenamefont {Khaliullin}(2013)}]{khaliullin2013excitonic}%
  \BibitemOpen
  \bibfield  {author} {\bibinfo {author} {\bibfnamefont {Giniyat}\ \bibnamefont
  {Khaliullin}},\ }\bibfield  {title} {\enquote {\bibinfo {title} {Excitonic
  magnetism in van vleck--type d 4 mott insulators},}\ }\href@noop {}
  {\bibfield  {journal} {\bibinfo  {journal} {Physical review letters}\
  }\textbf {\bibinfo {volume} {111}},\ \bibinfo {pages} {197201} (\bibinfo
  {year} {2013})}\BibitemShut {NoStop}%
\bibitem [{\citenamefont {Meetei}\ \emph {et~al.}(2015)\citenamefont {Meetei},
  \citenamefont {Cole}, \citenamefont {Randeria},\ and\ \citenamefont
  {Trivedi}}]{meetei2015novel}%
  \BibitemOpen
  \bibfield  {author} {\bibinfo {author} {\bibfnamefont {O~Nganba}\
  \bibnamefont {Meetei}}, \bibinfo {author} {\bibfnamefont {William~S}\
  \bibnamefont {Cole}}, \bibinfo {author} {\bibfnamefont {Mohit}\ \bibnamefont
  {Randeria}}, \ and\ \bibinfo {author} {\bibfnamefont {Nandini}\ \bibnamefont
  {Trivedi}},\ }\bibfield  {title} {\enquote {\bibinfo {title} {Novel magnetic
  state in d 4 mott insulators},}\ }\href@noop {} {\bibfield  {journal}
  {\bibinfo  {journal} {Physical Review B}\ }\textbf {\bibinfo {volume} {91}},\
  \bibinfo {pages} {054412} (\bibinfo {year} {2015})}\BibitemShut {NoStop}%
\bibitem [{\citenamefont {Cao}\ \emph {et~al.}(2014)\citenamefont {Cao},
  \citenamefont {Qi}, \citenamefont {Li}, \citenamefont {Terzic}, \citenamefont
  {Yuan}, \citenamefont {DeLong}, \citenamefont {Murthy},\ and\ \citenamefont
  {Kaul}}]{cao2014novel}%
  \BibitemOpen
  \bibfield  {author} {\bibinfo {author} {\bibfnamefont {Gang}\ \bibnamefont
  {Cao}}, \bibinfo {author} {\bibfnamefont {TF}~\bibnamefont {Qi}}, \bibinfo
  {author} {\bibfnamefont {Li}~\bibnamefont {Li}}, \bibinfo {author}
  {\bibfnamefont {Jsaminka}\ \bibnamefont {Terzic}}, \bibinfo {author}
  {\bibfnamefont {SJ}~\bibnamefont {Yuan}}, \bibinfo {author} {\bibfnamefont
  {Lance~E}\ \bibnamefont {DeLong}}, \bibinfo {author} {\bibfnamefont
  {Ganpathy}\ \bibnamefont {Murthy}}, \ and\ \bibinfo {author} {\bibfnamefont
  {Ribhu~K}\ \bibnamefont {Kaul}},\ }\bibfield  {title} {\enquote {\bibinfo
  {title} {Novel magnetism of ir 5+(5 d 4) ions in the double perovskite sr 2
  yiro 6},}\ }\href@noop {} {\bibfield  {journal} {\bibinfo  {journal}
  {Physical review letters}\ }\textbf {\bibinfo {volume} {112}},\ \bibinfo
  {pages} {056402} (\bibinfo {year} {2014})}\BibitemShut {NoStop}%
\bibitem [{\citenamefont {Nag}\ \emph {et~al.}(2016)\citenamefont {Nag},
  \citenamefont {Middey}, \citenamefont {Bhowal}, \citenamefont {Panda},
  \citenamefont {Mathieu}, \citenamefont {Orain}, \citenamefont {Bert},
  \citenamefont {Mendels}, \citenamefont {Freeman}, \citenamefont {Mansson}
  \emph {et~al.}}]{nag2016origin}%
  \BibitemOpen
  \bibfield  {author} {\bibinfo {author} {\bibfnamefont {Abhishek}\
  \bibnamefont {Nag}}, \bibinfo {author} {\bibfnamefont {Srimanta}\
  \bibnamefont {Middey}}, \bibinfo {author} {\bibfnamefont {Sayantika}\
  \bibnamefont {Bhowal}}, \bibinfo {author} {\bibfnamefont {Swarup~K}\
  \bibnamefont {Panda}}, \bibinfo {author} {\bibfnamefont {Roland}\
  \bibnamefont {Mathieu}}, \bibinfo {author} {\bibfnamefont {JC}~\bibnamefont
  {Orain}}, \bibinfo {author} {\bibfnamefont {F}~\bibnamefont {Bert}}, \bibinfo
  {author} {\bibfnamefont {P}~\bibnamefont {Mendels}}, \bibinfo {author}
  {\bibfnamefont {Paul~Gregory}\ \bibnamefont {Freeman}}, \bibinfo {author}
  {\bibfnamefont {M}~\bibnamefont {Mansson}},  \emph {et~al.},\ }\bibfield
  {title} {\enquote {\bibinfo {title} {Origin of the spin-orbital liquid state
  in a nearly j= 0 iridate ba 3 znir 2 o 9},}\ }\href@noop {} {\bibfield
  {journal} {\bibinfo  {journal} {Physical review letters}\ }\textbf {\bibinfo
  {volume} {116}},\ \bibinfo {pages} {097205} (\bibinfo {year}
  {2016})}\BibitemShut {NoStop}%
\bibitem [{\citenamefont {Dey}\ \emph {et~al.}(2016)\citenamefont {Dey},
  \citenamefont {Maljuk}, \citenamefont {Efremov}, \citenamefont {Kataeva},
  \citenamefont {Gass}, \citenamefont {Blum}, \citenamefont {Steckel},
  \citenamefont {Gruner}, \citenamefont {Ritschel}, \citenamefont {Wolter}
  \emph {et~al.}}]{dey2016ba}%
  \BibitemOpen
  \bibfield  {author} {\bibinfo {author} {\bibfnamefont {T}~\bibnamefont
  {Dey}}, \bibinfo {author} {\bibfnamefont {A}~\bibnamefont {Maljuk}}, \bibinfo
  {author} {\bibfnamefont {DV}~\bibnamefont {Efremov}}, \bibinfo {author}
  {\bibfnamefont {O}~\bibnamefont {Kataeva}}, \bibinfo {author} {\bibfnamefont
  {S}~\bibnamefont {Gass}}, \bibinfo {author} {\bibfnamefont {CGF}\
  \bibnamefont {Blum}}, \bibinfo {author} {\bibfnamefont {F}~\bibnamefont
  {Steckel}}, \bibinfo {author} {\bibfnamefont {D}~\bibnamefont {Gruner}},
  \bibinfo {author} {\bibfnamefont {T}~\bibnamefont {Ritschel}}, \bibinfo
  {author} {\bibfnamefont {AUB}\ \bibnamefont {Wolter}},  \emph {et~al.},\
  }\bibfield  {title} {\enquote {\bibinfo {title} {Ba 2 yiro 6: a cubic double
  perovskite material with ir 5+ ions},}\ }\href@noop {} {\bibfield  {journal}
  {\bibinfo  {journal} {Physical Review B}\ }\textbf {\bibinfo {volume} {93}},\
  \bibinfo {pages} {014434} (\bibinfo {year} {2016})}\BibitemShut {NoStop}%
\bibitem [{\citenamefont {Bhowal}\ \emph {et~al.}(2015)\citenamefont {Bhowal},
  \citenamefont {Baidya}, \citenamefont {Dasgupta},\ and\ \citenamefont
  {Saha-Dasgupta}}]{bhowal2015breakdown}%
  \BibitemOpen
  \bibfield  {author} {\bibinfo {author} {\bibfnamefont {Sayantika}\
  \bibnamefont {Bhowal}}, \bibinfo {author} {\bibfnamefont {Santu}\
  \bibnamefont {Baidya}}, \bibinfo {author} {\bibfnamefont {Indra}\
  \bibnamefont {Dasgupta}}, \ and\ \bibinfo {author} {\bibfnamefont {Tanusri}\
  \bibnamefont {Saha-Dasgupta}},\ }\bibfield  {title} {\enquote {\bibinfo
  {title} {Breakdown of j= 0 nonmagnetic state in d 4 iridate double
  perovskites: A first-principles study},}\ }\href@noop {} {\bibfield
  {journal} {\bibinfo  {journal} {Physical Review B}\ }\textbf {\bibinfo
  {volume} {92}},\ \bibinfo {pages} {121113} (\bibinfo {year}
  {2015})}\BibitemShut {NoStop}%
\bibitem [{\citenamefont {Corredor}\ \emph {et~al.}(2017)\citenamefont
  {Corredor}, \citenamefont {Aslan-Cansever}, \citenamefont {Sturza},
  \citenamefont {Manna}, \citenamefont {Maljuk}, \citenamefont {Gass},
  \citenamefont {Dey}, \citenamefont {Wolter}, \citenamefont {Kataeva},
  \citenamefont {Zimmermann} \emph {et~al.}}]{corredor2017iridium}%
  \BibitemOpen
  \bibfield  {author} {\bibinfo {author} {\bibfnamefont {LT}~\bibnamefont
  {Corredor}}, \bibinfo {author} {\bibfnamefont {G}~\bibnamefont
  {Aslan-Cansever}}, \bibinfo {author} {\bibfnamefont {M}~\bibnamefont
  {Sturza}}, \bibinfo {author} {\bibfnamefont {Kaustuv}\ \bibnamefont {Manna}},
  \bibinfo {author} {\bibfnamefont {A}~\bibnamefont {Maljuk}}, \bibinfo
  {author} {\bibfnamefont {S}~\bibnamefont {Gass}}, \bibinfo {author}
  {\bibfnamefont {T}~\bibnamefont {Dey}}, \bibinfo {author} {\bibfnamefont
  {AUB}\ \bibnamefont {Wolter}}, \bibinfo {author} {\bibfnamefont {Olga}\
  \bibnamefont {Kataeva}}, \bibinfo {author} {\bibfnamefont {A}~\bibnamefont
  {Zimmermann}},  \emph {et~al.},\ }\bibfield  {title} {\enquote {\bibinfo
  {title} {Iridium double perovskite sr 2 yiro 6: A combined structural and
  specific heat study},}\ }\href@noop {} {\bibfield  {journal} {\bibinfo
  {journal} {Physical Review B}\ }\textbf {\bibinfo {volume} {95}},\ \bibinfo
  {pages} {064418} (\bibinfo {year} {2017})}\BibitemShut {NoStop}%
\bibitem [{\citenamefont {Pajskr}\ \emph {et~al.}(2016)\citenamefont {Pajskr},
  \citenamefont {Nov{\'a}k}, \citenamefont {Pokorn{\`y}}, \citenamefont
  {Koloren{\v{c}}}, \citenamefont {Arita},\ and\ \citenamefont
  {Kune{\v{s}}}}]{pajskr2016possibility}%
  \BibitemOpen
  \bibfield  {author} {\bibinfo {author} {\bibfnamefont {K}~\bibnamefont
  {Pajskr}}, \bibinfo {author} {\bibfnamefont {P}~\bibnamefont {Nov{\'a}k}},
  \bibinfo {author} {\bibfnamefont {V}~\bibnamefont {Pokorn{\`y}}}, \bibinfo
  {author} {\bibfnamefont {J}~\bibnamefont {Koloren{\v{c}}}}, \bibinfo {author}
  {\bibfnamefont {R}~\bibnamefont {Arita}}, \ and\ \bibinfo {author}
  {\bibfnamefont {J}~\bibnamefont {Kune{\v{s}}}},\ }\bibfield  {title}
  {\enquote {\bibinfo {title} {On the possibility of excitonic magnetism in ir
  double perovskites},}\ }\href@noop {} {\bibfield  {journal} {\bibinfo
  {journal} {Physical Review B}\ }\textbf {\bibinfo {volume} {93}},\ \bibinfo
  {pages} {035129} (\bibinfo {year} {2016})}\BibitemShut {NoStop}%
\bibitem [{\citenamefont {Ranjbar}\ \emph {et~al.}(2015)\citenamefont
  {Ranjbar}, \citenamefont {Reynolds}, \citenamefont {Kayser}, \citenamefont
  {Kennedy}, \citenamefont {Hester},\ and\ \citenamefont
  {Kimpton}}]{ranjbar2015structural}%
  \BibitemOpen
  \bibfield  {author} {\bibinfo {author} {\bibfnamefont {Ben}\ \bibnamefont
  {Ranjbar}}, \bibinfo {author} {\bibfnamefont {Emily}\ \bibnamefont
  {Reynolds}}, \bibinfo {author} {\bibfnamefont {Paula}\ \bibnamefont
  {Kayser}}, \bibinfo {author} {\bibfnamefont {Brendan~J}\ \bibnamefont
  {Kennedy}}, \bibinfo {author} {\bibfnamefont {James~R}\ \bibnamefont
  {Hester}}, \ and\ \bibinfo {author} {\bibfnamefont {Justin~A}\ \bibnamefont
  {Kimpton}},\ }\bibfield  {title} {\enquote {\bibinfo {title} {Structural and
  magnetic properties of the iridium double perovskites ba2--x sr x yiro6},}\
  }\href@noop {} {\bibfield  {journal} {\bibinfo  {journal} {Inorganic
  chemistry}\ }\textbf {\bibinfo {volume} {54}},\ \bibinfo {pages}
  {10468--10476} (\bibinfo {year} {2015})}\BibitemShut {NoStop}%
\bibitem [{\citenamefont {Chaloupka}\ and\ \citenamefont
  {Khaliullin}(2016)}]{chaloupka2016doping}%
  \BibitemOpen
  \bibfield  {author} {\bibinfo {author} {\bibfnamefont {Ji{\v{r}}{\'\i}}\
  \bibnamefont {Chaloupka}}\ and\ \bibinfo {author} {\bibfnamefont {Giniyat}\
  \bibnamefont {Khaliullin}},\ }\bibfield  {title} {\enquote {\bibinfo {title}
  {Doping-induced ferromagnetism and possible triplet pairing in d 4 mott
  insulators},}\ }\href@noop {} {\bibfield  {journal} {\bibinfo  {journal}
  {Physical review letters}\ }\textbf {\bibinfo {volume} {116}},\ \bibinfo
  {pages} {017203} (\bibinfo {year} {2016})}\BibitemShut {NoStop}%
\bibitem [{\citenamefont {Terzic}\ \emph {et~al.}(2017)\citenamefont {Terzic},
  \citenamefont {Zheng}, \citenamefont {Ye}, \citenamefont {Zhao},
  \citenamefont {Schlottmann}, \citenamefont {De~Long}, \citenamefont {Yuan},\
  and\ \citenamefont {Cao}}]{terzic2017evidence}%
  \BibitemOpen
  \bibfield  {author} {\bibinfo {author} {\bibfnamefont {Jasminka}\
  \bibnamefont {Terzic}}, \bibinfo {author} {\bibfnamefont {Hao}\ \bibnamefont
  {Zheng}}, \bibinfo {author} {\bibfnamefont {Feng}\ \bibnamefont {Ye}},
  \bibinfo {author} {\bibfnamefont {HD}~\bibnamefont {Zhao}}, \bibinfo {author}
  {\bibfnamefont {P}~\bibnamefont {Schlottmann}}, \bibinfo {author}
  {\bibfnamefont {Lance~E}\ \bibnamefont {De~Long}}, \bibinfo {author}
  {\bibfnamefont {SJ}~\bibnamefont {Yuan}}, \ and\ \bibinfo {author}
  {\bibfnamefont {Gang}\ \bibnamefont {Cao}},\ }\bibfield  {title} {\enquote
  {\bibinfo {title} {Evidence for a low-temperature magnetic ground state in
  double-perovskite iridates with i r 5+(5 d 4) ions},}\ }\href@noop {}
  {\bibfield  {journal} {\bibinfo  {journal} {Physical Review B}\ }\textbf
  {\bibinfo {volume} {96}},\ \bibinfo {pages} {064436} (\bibinfo {year}
  {2017})}\BibitemShut {NoStop}%
\bibitem [{\citenamefont {Phelan}\ \emph {et~al.}(2016)\citenamefont {Phelan},
  \citenamefont {Seibel}, \citenamefont {Badoe~Jr}, \citenamefont {Xie},\ and\
  \citenamefont {Cava}}]{phelan2016influence}%
  \BibitemOpen
  \bibfield  {author} {\bibinfo {author} {\bibfnamefont {Brendan~F}\
  \bibnamefont {Phelan}}, \bibinfo {author} {\bibfnamefont {Elizabeth~M}\
  \bibnamefont {Seibel}}, \bibinfo {author} {\bibfnamefont {Daniel}\
  \bibnamefont {Badoe~Jr}}, \bibinfo {author} {\bibfnamefont {Weiwei}\
  \bibnamefont {Xie}}, \ and\ \bibinfo {author} {\bibfnamefont
  {RJ}~\bibnamefont {Cava}},\ }\bibfield  {title} {\enquote {\bibinfo {title}
  {Influence of structural distortions on the ir magnetism in ba2- xsrxyiro6
  double perovskites},}\ }\href@noop {} {\bibfield  {journal} {\bibinfo
  {journal} {Solid State Communications}\ }\textbf {\bibinfo {volume} {236}},\
  \bibinfo {pages} {37--40} (\bibinfo {year} {2016})}\BibitemShut {NoStop}%
\bibitem [{\citenamefont {Fuchs}\ \emph {et~al.}(2018)\citenamefont {Fuchs},
  \citenamefont {Dey}, \citenamefont {Aslan-Cansever}, \citenamefont {Maljuk},
  \citenamefont {Wurmehl}, \citenamefont {B{\"u}chner},\ and\ \citenamefont
  {Kataev}}]{fuchs2018unraveling}%
  \BibitemOpen
  \bibfield  {author} {\bibinfo {author} {\bibfnamefont {S}~\bibnamefont
  {Fuchs}}, \bibinfo {author} {\bibfnamefont {T}~\bibnamefont {Dey}}, \bibinfo
  {author} {\bibfnamefont {G}~\bibnamefont {Aslan-Cansever}}, \bibinfo {author}
  {\bibfnamefont {A}~\bibnamefont {Maljuk}}, \bibinfo {author} {\bibfnamefont
  {S}~\bibnamefont {Wurmehl}}, \bibinfo {author} {\bibfnamefont
  {B}~\bibnamefont {B{\"u}chner}}, \ and\ \bibinfo {author} {\bibfnamefont
  {V}~\bibnamefont {Kataev}},\ }\bibfield  {title} {\enquote {\bibinfo {title}
  {Unraveling the nature of magnetism of the 5 d 4 double perovskite ba 2 yiro
  6},}\ }\href@noop {} {\bibfield  {journal} {\bibinfo  {journal} {Physical
  review letters}\ }\textbf {\bibinfo {volume} {120}},\ \bibinfo {pages}
  {237204} (\bibinfo {year} {2018})}\BibitemShut {NoStop}%
\bibitem [{\citenamefont {Chen}\ \emph {et~al.}(2017)\citenamefont {Chen},
  \citenamefont {Svoboda}, \citenamefont {Zheng}, \citenamefont {Sales},
  \citenamefont {Mandrus}, \citenamefont {Zhou}, \citenamefont {Zhou},
  \citenamefont {McComb}, \citenamefont {Randeria}, \citenamefont {Trivedi}
  \emph {et~al.}}]{chen2017magnetism}%
  \BibitemOpen
  \bibfield  {author} {\bibinfo {author} {\bibfnamefont {Qiang}\ \bibnamefont
  {Chen}}, \bibinfo {author} {\bibfnamefont {Chris}\ \bibnamefont {Svoboda}},
  \bibinfo {author} {\bibfnamefont {Qiang}\ \bibnamefont {Zheng}}, \bibinfo
  {author} {\bibfnamefont {Brian~C}\ \bibnamefont {Sales}}, \bibinfo {author}
  {\bibfnamefont {David~G}\ \bibnamefont {Mandrus}}, \bibinfo {author}
  {\bibfnamefont {HD}~\bibnamefont {Zhou}}, \bibinfo {author} {\bibfnamefont
  {J-S}\ \bibnamefont {Zhou}}, \bibinfo {author} {\bibfnamefont
  {D}~\bibnamefont {McComb}}, \bibinfo {author} {\bibfnamefont {Mohit}\
  \bibnamefont {Randeria}}, \bibinfo {author} {\bibfnamefont {Nandini}\
  \bibnamefont {Trivedi}},  \emph {et~al.},\ }\bibfield  {title} {\enquote
  {\bibinfo {title} {Magnetism out of antisite disorder in the j= 0 compound ba
  2 yiro 6},}\ }\href@noop {} {\bibfield  {journal} {\bibinfo  {journal}
  {Physical Review B}\ }\textbf {\bibinfo {volume} {96}},\ \bibinfo {pages}
  {144423} (\bibinfo {year} {2017})}\BibitemShut {NoStop}%
\bibitem [{\citenamefont {Hammerath}\ \emph {et~al.}(2017)\citenamefont
  {Hammerath}, \citenamefont {Sarkar}, \citenamefont {Kamusella}, \citenamefont
  {Baines}, \citenamefont {Klauss}, \citenamefont {Dey}, \citenamefont
  {Maljuk}, \citenamefont {Ga{\ss}}, \citenamefont {Wolter}, \citenamefont
  {Grafe} \emph {et~al.}}]{hammerath2017diluted}%
  \BibitemOpen
  \bibfield  {author} {\bibinfo {author} {\bibfnamefont {F}~\bibnamefont
  {Hammerath}}, \bibinfo {author} {\bibfnamefont {R}~\bibnamefont {Sarkar}},
  \bibinfo {author} {\bibfnamefont {S}~\bibnamefont {Kamusella}}, \bibinfo
  {author} {\bibfnamefont {C}~\bibnamefont {Baines}}, \bibinfo {author}
  {\bibfnamefont {H-H}\ \bibnamefont {Klauss}}, \bibinfo {author}
  {\bibfnamefont {T}~\bibnamefont {Dey}}, \bibinfo {author} {\bibfnamefont
  {A}~\bibnamefont {Maljuk}}, \bibinfo {author} {\bibfnamefont {S}~\bibnamefont
  {Ga{\ss}}}, \bibinfo {author} {\bibfnamefont {AUB}\ \bibnamefont {Wolter}},
  \bibinfo {author} {\bibfnamefont {H-J}\ \bibnamefont {Grafe}},  \emph
  {et~al.},\ }\bibfield  {title} {\enquote {\bibinfo {title} {Diluted
  paramagnetic impurities in nonmagnetic ba 2 yiro 6},}\ }\href@noop {}
  {\bibfield  {journal} {\bibinfo  {journal} {Physical Review B}\ }\textbf
  {\bibinfo {volume} {96}},\ \bibinfo {pages} {165108} (\bibinfo {year}
  {2017})}\BibitemShut {NoStop}%
\bibitem [{\citenamefont {Zhao}\ \emph {et~al.}(2016)\citenamefont {Zhao},
  \citenamefont {Calder}, \citenamefont {Aczel}, \citenamefont {McGuire},
  \citenamefont {Sales}, \citenamefont {Mandrus}, \citenamefont {Chen},
  \citenamefont {Trivedi}, \citenamefont {Zhou},\ and\ \citenamefont
  {Yan}}]{zhao2016fragile}%
  \BibitemOpen
  \bibfield  {author} {\bibinfo {author} {\bibfnamefont {ZY}~\bibnamefont
  {Zhao}}, \bibinfo {author} {\bibfnamefont {S}~\bibnamefont {Calder}},
  \bibinfo {author} {\bibfnamefont {AA}~\bibnamefont {Aczel}}, \bibinfo
  {author} {\bibfnamefont {MA}~\bibnamefont {McGuire}}, \bibinfo {author}
  {\bibfnamefont {BC}~\bibnamefont {Sales}}, \bibinfo {author} {\bibfnamefont
  {DG}~\bibnamefont {Mandrus}}, \bibinfo {author} {\bibfnamefont
  {G}~\bibnamefont {Chen}}, \bibinfo {author} {\bibfnamefont {N}~\bibnamefont
  {Trivedi}}, \bibinfo {author} {\bibfnamefont {HD}~\bibnamefont {Zhou}}, \
  and\ \bibinfo {author} {\bibfnamefont {J-Q}\ \bibnamefont {Yan}},\ }\bibfield
   {title} {\enquote {\bibinfo {title} {Fragile singlet ground-state magnetism
  in the pyrochlore osmates r 2 os 2 o 7 (r= y and ho)},}\ }\href@noop {}
  {\bibfield  {journal} {\bibinfo  {journal} {Physical Review B}\ }\textbf
  {\bibinfo {volume} {93}},\ \bibinfo {pages} {134426} (\bibinfo {year}
  {2016})}\BibitemShut {NoStop}%
\bibitem [{\citenamefont {Wolff}\ \emph {et~al.}(2017)\citenamefont {Wolff},
  \citenamefont {Agrestini}, \citenamefont {Tanaka}, \citenamefont {Jansen},\
  and\ \citenamefont {Tjeng}}]{wolff2017comparative}%
  \BibitemOpen
  \bibfield  {author} {\bibinfo {author} {\bibfnamefont {Klaus~K}\ \bibnamefont
  {Wolff}}, \bibinfo {author} {\bibfnamefont {Stefano}\ \bibnamefont
  {Agrestini}}, \bibinfo {author} {\bibfnamefont {Arata}\ \bibnamefont
  {Tanaka}}, \bibinfo {author} {\bibfnamefont {Martin}\ \bibnamefont {Jansen}},
  \ and\ \bibinfo {author} {\bibfnamefont {Liu~Hao}\ \bibnamefont {Tjeng}},\
  }\bibfield  {title} {\enquote {\bibinfo {title} {Comparative study of
  potentially jeff= 0 ground state iridium (v) in srlaniiro6, srlamgiro6, and
  srlazniro6},}\ }\href@noop {} {\bibfield  {journal} {\bibinfo  {journal}
  {Zeitschrift f{\"u}r anorganische und allgemeine Chemie}\ }\textbf {\bibinfo
  {volume} {643}},\ \bibinfo {pages} {2095--2101} (\bibinfo {year}
  {2017})}\BibitemShut {NoStop}%
\bibitem [{\citenamefont {Prasad}\ \emph {et~al.}(2018)\citenamefont {Prasad},
  \citenamefont {Doert}, \citenamefont {Felser},\ and\ \citenamefont
  {Jansen}}]{prasad2018jeff}%
  \BibitemOpen
  \bibfield  {author} {\bibinfo {author} {\bibfnamefont {Beluvalli~E}\
  \bibnamefont {Prasad}}, \bibinfo {author} {\bibfnamefont {Thomas}\
  \bibnamefont {Doert}}, \bibinfo {author} {\bibfnamefont {Claudia}\
  \bibnamefont {Felser}}, \ and\ \bibinfo {author} {\bibfnamefont {Martin}\
  \bibnamefont {Jansen}},\ }\bibfield  {title} {\enquote {\bibinfo {title} {On
  jeff= 0 ground state iridates (v): Tracking residual paramagnetism in new
  bi2nairo6},}\ }\href@noop {} {\bibfield  {journal} {\bibinfo  {journal}
  {Chemistry--A European Journal}\ }\textbf {\bibinfo {volume} {24}},\ \bibinfo
  {pages} {16762--16765} (\bibinfo {year} {2018})}\BibitemShut {NoStop}%
\bibitem [{\citenamefont {Laguna-Marco}\ \emph {et~al.}(2020)\citenamefont
  {Laguna-Marco}, \citenamefont {Arias-Egido}, \citenamefont {Piquer},
  \citenamefont {Cuartero}, \citenamefont {Hern{\'a}ndez-L{\'o}pez},
  \citenamefont {Kayser}, \citenamefont {Alonso}, \citenamefont {Barker},
  \citenamefont {Fabbris}, \citenamefont {Escanhoela~Jr} \emph
  {et~al.}}]{laguna2020magnetism}%
  \BibitemOpen
  \bibfield  {author} {\bibinfo {author} {\bibfnamefont {MA}~\bibnamefont
  {Laguna-Marco}}, \bibinfo {author} {\bibfnamefont {E}~\bibnamefont
  {Arias-Egido}}, \bibinfo {author} {\bibfnamefont {Cristina}\ \bibnamefont
  {Piquer}}, \bibinfo {author} {\bibfnamefont {Vera}\ \bibnamefont {Cuartero}},
  \bibinfo {author} {\bibfnamefont {Leyre}\ \bibnamefont
  {Hern{\'a}ndez-L{\'o}pez}}, \bibinfo {author} {\bibfnamefont {Paula}\
  \bibnamefont {Kayser}}, \bibinfo {author} {\bibfnamefont {JA}~\bibnamefont
  {Alonso}}, \bibinfo {author} {\bibfnamefont {JAT}\ \bibnamefont {Barker}},
  \bibinfo {author} {\bibfnamefont {G}~\bibnamefont {Fabbris}}, \bibinfo
  {author} {\bibfnamefont {CA}~\bibnamefont {Escanhoela~Jr}},  \emph {et~al.},\
  }\bibfield  {title} {\enquote {\bibinfo {title} {Magnetism of ir 5+-based
  double perovskites: Unraveling its nature and the influence of structure},}\
  }\href@noop {} {\bibfield  {journal} {\bibinfo  {journal} {Physical Review
  B}\ }\textbf {\bibinfo {volume} {101}},\ \bibinfo {pages} {014449} (\bibinfo
  {year} {2020})}\BibitemShut {NoStop}%
\bibitem [{\citenamefont {Davies}\ \emph {et~al.}(2019)\citenamefont {Davies},
  \citenamefont {Topping}, \citenamefont {Jacobsen}, \citenamefont {Princep},
  \citenamefont {Kirschner}, \citenamefont {Rahn}, \citenamefont {Bristow},
  \citenamefont {Vale}, \citenamefont {Da~Silva}, \citenamefont {Baker} \emph
  {et~al.}}]{davies2019evidence}%
  \BibitemOpen
  \bibfield  {author} {\bibinfo {author} {\bibfnamefont {NR}~\bibnamefont
  {Davies}}, \bibinfo {author} {\bibfnamefont {CV}~\bibnamefont {Topping}},
  \bibinfo {author} {\bibfnamefont {H}~\bibnamefont {Jacobsen}}, \bibinfo
  {author} {\bibfnamefont {AJ}~\bibnamefont {Princep}}, \bibinfo {author}
  {\bibfnamefont {FKK}\ \bibnamefont {Kirschner}}, \bibinfo {author}
  {\bibfnamefont {MC}~\bibnamefont {Rahn}}, \bibinfo {author} {\bibfnamefont
  {M}~\bibnamefont {Bristow}}, \bibinfo {author} {\bibfnamefont
  {JG}~\bibnamefont {Vale}}, \bibinfo {author} {\bibfnamefont {I}~\bibnamefont
  {Da~Silva}}, \bibinfo {author} {\bibfnamefont {PJ}~\bibnamefont {Baker}},
  \emph {et~al.},\ }\bibfield  {title} {\enquote {\bibinfo {title} {Evidence
  for a j eff= 0 ground state and defect-induced spin glass behavior in the
  pyrochlore osmate y 2 os 2 o 7},}\ }\href@noop {} {\bibfield  {journal}
  {\bibinfo  {journal} {Physical Review B}\ }\textbf {\bibinfo {volume} {99}},\
  \bibinfo {pages} {174442} (\bibinfo {year} {2019})}\BibitemShut {NoStop}%
\bibitem [{\citenamefont {Gong}\ \emph {et~al.}(2018)\citenamefont {Gong},
  \citenamefont {Kim}, \citenamefont {Kim}, \citenamefont {Kim}, \citenamefont
  {Kim},\ and\ \citenamefont {Min}}]{gong2018ground}%
  \BibitemOpen
  \bibfield  {author} {\bibinfo {author} {\bibfnamefont {Hoshin}\ \bibnamefont
  {Gong}}, \bibinfo {author} {\bibfnamefont {Kyoo}\ \bibnamefont {Kim}},
  \bibinfo {author} {\bibfnamefont {Beom~Hyun}\ \bibnamefont {Kim}}, \bibinfo
  {author} {\bibfnamefont {Bongjae}\ \bibnamefont {Kim}}, \bibinfo {author}
  {\bibfnamefont {Junwon}\ \bibnamefont {Kim}}, \ and\ \bibinfo {author}
  {\bibfnamefont {BI}~\bibnamefont {Min}},\ }\bibfield  {title} {\enquote
  {\bibinfo {title} {Is the ground state of 5d4 double-perovskite iridate
  ba2yiro6 magnetic or nonmagnetic?}}\ }\href@noop {} {\bibfield  {journal}
  {\bibinfo  {journal} {Journal of Magnetism and Magnetic Materials}\ }\textbf
  {\bibinfo {volume} {454}},\ \bibinfo {pages} {66--70} (\bibinfo {year}
  {2018})}\BibitemShut {NoStop}%
\bibitem [{\citenamefont {Long}\ and\ \citenamefont
  {Shimakawa}(2010)}]{long2010intermetallic}%
  \BibitemOpen
  \bibfield  {author} {\bibinfo {author} {\bibfnamefont {Youwen}\ \bibnamefont
  {Long}}\ and\ \bibinfo {author} {\bibfnamefont {Yuichi}\ \bibnamefont
  {Shimakawa}},\ }\bibfield  {title} {\enquote {\bibinfo {title} {Intermetallic
  charge transfer between a-site cu and b-site fe in a-site-ordered double
  perovskites},}\ }\href@noop {} {\bibfield  {journal} {\bibinfo  {journal}
  {New Journal of Physics}\ }\textbf {\bibinfo {volume} {12}},\ \bibinfo
  {pages} {063029} (\bibinfo {year} {2010})}\BibitemShut {NoStop}%
\bibitem [{\citenamefont {Hejtm{\'a}nek}\ \emph {et~al.}(2010)\citenamefont
  {Hejtm{\'a}nek}, \citenamefont {{\v{S}}antav{\'a}}, \citenamefont
  {Kn{\'\i}{\v{z}}ek}, \citenamefont {Mary{\v{s}}ko}, \citenamefont
  {Jir{\'a}k}, \citenamefont {Naito}, \citenamefont {Sasaki},\ and\
  \citenamefont {Fujishiro}}]{hejtmanek2010metal}%
  \BibitemOpen
  \bibfield  {author} {\bibinfo {author} {\bibfnamefont {J}~\bibnamefont
  {Hejtm{\'a}nek}}, \bibinfo {author} {\bibfnamefont {E}~\bibnamefont
  {{\v{S}}antav{\'a}}}, \bibinfo {author} {\bibfnamefont {K}~\bibnamefont
  {Kn{\'\i}{\v{z}}ek}}, \bibinfo {author} {\bibfnamefont {M}~\bibnamefont
  {Mary{\v{s}}ko}}, \bibinfo {author} {\bibfnamefont {Z}~\bibnamefont
  {Jir{\'a}k}}, \bibinfo {author} {\bibfnamefont {T}~\bibnamefont {Naito}},
  \bibinfo {author} {\bibfnamefont {H}~\bibnamefont {Sasaki}}, \ and\ \bibinfo
  {author} {\bibfnamefont {H}~\bibnamefont {Fujishiro}},\ }\bibfield  {title}
  {\enquote {\bibinfo {title} {Metal-insulator transition and the pr 3+/pr 4+
  valence shift in (pr 1- y y y) 0.7 ca 0.3 coo 3},}\ }\href@noop {} {\bibfield
   {journal} {\bibinfo  {journal} {Physical Review B}\ }\textbf {\bibinfo
  {volume} {82}},\ \bibinfo {pages} {165107} (\bibinfo {year}
  {2010})}\BibitemShut {NoStop}%
\bibitem [{\citenamefont {Zhou}\ \emph {et~al.}(2005)\citenamefont {Zhou},
  \citenamefont {Kennedy}, \citenamefont {Wallwork}, \citenamefont {Elcombe},
  \citenamefont {Lee},\ and\ \citenamefont {Vogt}}]{zhou2005temperature}%
  \BibitemOpen
  \bibfield  {author} {\bibinfo {author} {\bibfnamefont {Qingdi}\ \bibnamefont
  {Zhou}}, \bibinfo {author} {\bibfnamefont {Brendan~J}\ \bibnamefont
  {Kennedy}}, \bibinfo {author} {\bibfnamefont {Kia~S}\ \bibnamefont
  {Wallwork}}, \bibinfo {author} {\bibfnamefont {Margaret~M}\ \bibnamefont
  {Elcombe}}, \bibinfo {author} {\bibfnamefont {Yongjae}\ \bibnamefont {Lee}},
  \ and\ \bibinfo {author} {\bibfnamefont {Thomas}\ \bibnamefont {Vogt}},\
  }\bibfield  {title} {\enquote {\bibinfo {title} {Temperature and pressure
  dependent structural studies of the ordered double perovskites sr2tbru1-
  xirxo6},}\ }\href@noop {} {\bibfield  {journal} {\bibinfo  {journal} {Journal
  of Solid State Chemistry}\ }\textbf {\bibinfo {volume} {178}},\ \bibinfo
  {pages} {2282--2291} (\bibinfo {year} {2005})}\BibitemShut {NoStop}%
\bibitem [{\citenamefont {Doi}\ \emph {et~al.}(2000)\citenamefont {Doi},
  \citenamefont {Hinatsu}, \citenamefont {Oikawa}, \citenamefont {Shimojo},\
  and\ \citenamefont {Morii}}]{doi2000magnetic}%
  \BibitemOpen
  \bibfield  {author} {\bibinfo {author} {\bibfnamefont {Yoshihiro}\
  \bibnamefont {Doi}}, \bibinfo {author} {\bibfnamefont {Yukio}\ \bibnamefont
  {Hinatsu}}, \bibinfo {author} {\bibfnamefont {Ken-ichi}\ \bibnamefont
  {Oikawa}}, \bibinfo {author} {\bibfnamefont {Yutaka}\ \bibnamefont
  {Shimojo}}, \ and\ \bibinfo {author} {\bibfnamefont {Yukio}\ \bibnamefont
  {Morii}},\ }\bibfield  {title} {\enquote {\bibinfo {title} {Magnetic
  properties and crystal structures of ordered perovskites sr2tbru1- xirxo6},}\
  }\href@noop {} {\bibfield  {journal} {\bibinfo  {journal} {Journal of
  Materials Chemistry}\ }\textbf {\bibinfo {volume} {10}},\ \bibinfo {pages}
  {1731--1735} (\bibinfo {year} {2000})}\BibitemShut {NoStop}%
\bibitem [{\citenamefont {Wakeshima}\ \emph {et~al.}(2001)\citenamefont
  {Wakeshima}, \citenamefont {Izumiyama}, \citenamefont {Doi},\ and\
  \citenamefont {Hinatsu}}]{wakeshima2001valence}%
  \BibitemOpen
  \bibfield  {author} {\bibinfo {author} {\bibfnamefont {Makoto}\ \bibnamefont
  {Wakeshima}}, \bibinfo {author} {\bibfnamefont {Yuki}\ \bibnamefont
  {Izumiyama}}, \bibinfo {author} {\bibfnamefont {Yoshihiro}\ \bibnamefont
  {Doi}}, \ and\ \bibinfo {author} {\bibfnamefont {Yukio}\ \bibnamefont
  {Hinatsu}},\ }\bibfield  {title} {\enquote {\bibinfo {title} {Valence
  transition in ordered perovskites ba2prru1- xirxo6},}\ }\href@noop {}
  {\bibfield  {journal} {\bibinfo  {journal} {Solid state communications}\
  }\textbf {\bibinfo {volume} {120}},\ \bibinfo {pages} {273--278} (\bibinfo
  {year} {2001})}\BibitemShut {NoStop}%
\bibitem [{\citenamefont {Sannigrahi}\ \emph {et~al.}(2019)\citenamefont
  {Sannigrahi}, \citenamefont {Adroja}, \citenamefont {Ritter}, \citenamefont
  {Kockelmann}, \citenamefont {Hillier}, \citenamefont {Knight}, \citenamefont
  {Boothroyd}, \citenamefont {Wakeshima}, \citenamefont {Hinatsu},
  \citenamefont {Mosselmans} \emph {et~al.}}]{sannigrahi2019first}%
  \BibitemOpen
  \bibfield  {author} {\bibinfo {author} {\bibfnamefont {J}~\bibnamefont
  {Sannigrahi}}, \bibinfo {author} {\bibfnamefont {DT}~\bibnamefont {Adroja}},
  \bibinfo {author} {\bibfnamefont {C}~\bibnamefont {Ritter}}, \bibinfo
  {author} {\bibfnamefont {W}~\bibnamefont {Kockelmann}}, \bibinfo {author}
  {\bibfnamefont {AD}~\bibnamefont {Hillier}}, \bibinfo {author} {\bibfnamefont
  {KS}~\bibnamefont {Knight}}, \bibinfo {author} {\bibfnamefont
  {AT}~\bibnamefont {Boothroyd}}, \bibinfo {author} {\bibfnamefont
  {M}~\bibnamefont {Wakeshima}}, \bibinfo {author} {\bibfnamefont
  {Y}~\bibnamefont {Hinatsu}}, \bibinfo {author} {\bibfnamefont {JFW}\
  \bibnamefont {Mosselmans}},  \emph {et~al.},\ }\bibfield  {title} {\enquote
  {\bibinfo {title} {First-order valence transition: Neutron diffraction,
  inelastic neutron scattering, and x-ray absorption investigations on the
  double perovskite ba 2 prru 0.9 ir 0.1 o 6},}\ }\href@noop {} {\bibfield
  {journal} {\bibinfo  {journal} {Physical Review B}\ }\textbf {\bibinfo
  {volume} {99}},\ \bibinfo {pages} {184440} (\bibinfo {year}
  {2019})}\BibitemShut {NoStop}%
\bibitem [{\citenamefont {Zhou}\ and\ \citenamefont
  {Kennedy}(2005)}]{zhou2005independent}%
  \BibitemOpen
  \bibfield  {author} {\bibinfo {author} {\bibfnamefont {Qingdi}\ \bibnamefont
  {Zhou}}\ and\ \bibinfo {author} {\bibfnamefont {Brendan~J}\ \bibnamefont
  {Kennedy}},\ }\bibfield  {title} {\enquote {\bibinfo {title} {Independent
  structural and valence state transitions in the cation-ordered double
  perovskites ba2- xsrxtbiro6},}\ }\href@noop {} {\bibfield  {journal}
  {\bibinfo  {journal} {Journal of Solid State Chemistry}\ }\textbf {\bibinfo
  {volume} {178}},\ \bibinfo {pages} {3589--3594} (\bibinfo {year}
  {2005})}\BibitemShut {NoStop}%
\bibitem [{\citenamefont {Wakeshima}\ \emph {et~al.}(2000)\citenamefont
  {Wakeshima}, \citenamefont {Harada},\ and\ \citenamefont
  {Hinatsu}}]{wakeshima2000crystal}%
  \BibitemOpen
  \bibfield  {author} {\bibinfo {author} {\bibfnamefont {Makoto}\ \bibnamefont
  {Wakeshima}}, \bibinfo {author} {\bibfnamefont {Daijitsu}\ \bibnamefont
  {Harada}}, \ and\ \bibinfo {author} {\bibfnamefont {Yukio}\ \bibnamefont
  {Hinatsu}},\ }\bibfield  {title} {\enquote {\bibinfo {title} {Crystal
  structures and magnetic properties of ordered perovskites ba2lniro6 (ln=
  lanthanide)},}\ }\href@noop {} {\bibfield  {journal} {\bibinfo  {journal}
  {Journal of Materials Chemistry}\ }\textbf {\bibinfo {volume} {10}},\
  \bibinfo {pages} {419--422} (\bibinfo {year} {2000})}\BibitemShut {NoStop}%
\bibitem [{\citenamefont {Harada}\ \emph {et~al.}(2000)\citenamefont {Harada},
  \citenamefont {Wakeshima}, \citenamefont {Hinatsu}, \citenamefont {Ohoyama},\
  and\ \citenamefont {Yamaguchi}}]{harada2000magnetic}%
  \BibitemOpen
  \bibfield  {author} {\bibinfo {author} {\bibfnamefont {Daijitsu}\
  \bibnamefont {Harada}}, \bibinfo {author} {\bibfnamefont {Makoto}\
  \bibnamefont {Wakeshima}}, \bibinfo {author} {\bibfnamefont {Yukio}\
  \bibnamefont {Hinatsu}}, \bibinfo {author} {\bibfnamefont {Kenji}\
  \bibnamefont {Ohoyama}}, \ and\ \bibinfo {author} {\bibfnamefont {Yasuo}\
  \bibnamefont {Yamaguchi}},\ }\bibfield  {title} {\enquote {\bibinfo {title}
  {Magnetic and neutron diffraction study on iridium (iv) perovskites sr2lniro6
  (ln= ce, tb)},}\ }\href@noop {} {\bibfield  {journal} {\bibinfo  {journal}
  {Journal of Physics: Condensed Matter}\ }\textbf {\bibinfo {volume} {12}},\
  \bibinfo {pages} {3229} (\bibinfo {year} {2000})}\BibitemShut {NoStop}%
\bibitem [{\citenamefont {Harada}\ \emph {et~al.}(1999)\citenamefont {Harada},
  \citenamefont {Wakeshima},\ and\ \citenamefont
  {Hinatsu}}]{harada1999structure}%
  \BibitemOpen
  \bibfield  {author} {\bibinfo {author} {\bibfnamefont {Daijitsu}\
  \bibnamefont {Harada}}, \bibinfo {author} {\bibfnamefont {Makoto}\
  \bibnamefont {Wakeshima}}, \ and\ \bibinfo {author} {\bibfnamefont {Yukio}\
  \bibnamefont {Hinatsu}},\ }\bibfield  {title} {\enquote {\bibinfo {title}
  {The structure and magnetic properties of new iridium (iv) perovskites
  sr2lniro6 (ln= ce, tb)},}\ }\href@noop {} {\bibfield  {journal} {\bibinfo
  {journal} {Journal of Solid State Chemistry}\ }\textbf {\bibinfo {volume}
  {145}},\ \bibinfo {pages} {356--360} (\bibinfo {year} {1999})}\BibitemShut
  {NoStop}%
\bibitem [{\citenamefont {Ament}\ \emph {et~al.}(2011)\citenamefont {Ament},
  \citenamefont {Van~Veenendaal}, \citenamefont {Devereaux}, \citenamefont
  {Hill},\ and\ \citenamefont {Van Den~Brink}}]{ament2011resonant}%
  \BibitemOpen
  \bibfield  {author} {\bibinfo {author} {\bibfnamefont {Luuk~JP}\ \bibnamefont
  {Ament}}, \bibinfo {author} {\bibfnamefont {Michel}\ \bibnamefont
  {Van~Veenendaal}}, \bibinfo {author} {\bibfnamefont {Thomas~P}\ \bibnamefont
  {Devereaux}}, \bibinfo {author} {\bibfnamefont {John~P}\ \bibnamefont
  {Hill}}, \ and\ \bibinfo {author} {\bibfnamefont {Jeroen}\ \bibnamefont {Van
  Den~Brink}},\ }\bibfield  {title} {\enquote {\bibinfo {title} {Resonant
  inelastic x-ray scattering studies of elementary excitations},}\ }\href@noop
  {} {\bibfield  {journal} {\bibinfo  {journal} {Reviews of Modern Physics}\
  }\textbf {\bibinfo {volume} {83}},\ \bibinfo {pages} {705} (\bibinfo {year}
  {2011})}\BibitemShut {NoStop}%
\bibitem [{\citenamefont {Brese}\ and\ \citenamefont
  {O'keeffe}(1991)}]{brese1991bond}%
  \BibitemOpen
  \bibfield  {author} {\bibinfo {author} {\bibfnamefont {NE}~\bibnamefont
  {Brese}}\ and\ \bibinfo {author} {\bibfnamefont {M}~\bibnamefont
  {O'keeffe}},\ }\bibfield  {title} {\enquote {\bibinfo {title} {Bond-valence
  parameters for solids},}\ }\href@noop {} {\bibfield  {journal} {\bibinfo
  {journal} {Acta Crystallographica Section B: Structural Science}\ }\textbf
  {\bibinfo {volume} {47}},\ \bibinfo {pages} {192--197} (\bibinfo {year}
  {1991})}\BibitemShut {NoStop}%
\bibitem [{\citenamefont {Wills}(2000)}]{wills2000new}%
  \BibitemOpen
  \bibfield  {author} {\bibinfo {author} {\bibfnamefont {AS}~\bibnamefont
  {Wills}},\ }\bibfield  {title} {\enquote {\bibinfo {title} {A new protocol
  for the determination of magnetic structures using simulated annealing and
  representational analysis (sarah)},}\ }\href@noop {} {\bibfield  {journal}
  {\bibinfo  {journal} {Physica B: Condensed Matter}\ }\textbf {\bibinfo
  {volume} {276}},\ \bibinfo {pages} {680--681} (\bibinfo {year}
  {2000})}\BibitemShut {NoStop}%
\bibitem [{\citenamefont {Veber}\ \emph {et~al.}(2015)\citenamefont {Veber},
  \citenamefont {Vel{\'a}zquez}, \citenamefont {Gadret}, \citenamefont {Rytz},
  \citenamefont {Peltz},\ and\ \citenamefont {Decourt}}]{veber2015flux}%
  \BibitemOpen
  \bibfield  {author} {\bibinfo {author} {\bibfnamefont {Philippe}\
  \bibnamefont {Veber}}, \bibinfo {author} {\bibfnamefont {Matias}\
  \bibnamefont {Vel{\'a}zquez}}, \bibinfo {author} {\bibfnamefont
  {Gr{\'e}gory}\ \bibnamefont {Gadret}}, \bibinfo {author} {\bibfnamefont
  {Daniel}\ \bibnamefont {Rytz}}, \bibinfo {author} {\bibfnamefont {Mark}\
  \bibnamefont {Peltz}}, \ and\ \bibinfo {author} {\bibfnamefont {Rodolphe}\
  \bibnamefont {Decourt}},\ }\bibfield  {title} {\enquote {\bibinfo {title}
  {Flux growth at 1230 c of cubic tb 2 o 3 single crystals and characterization
  of their optical and magnetic properties},}\ }\href@noop {} {\bibfield
  {journal} {\bibinfo  {journal} {CrystEngComm}\ }\textbf {\bibinfo {volume}
  {17}},\ \bibinfo {pages} {492--497} (\bibinfo {year} {2015})}\BibitemShut
  {NoStop}%
\end{thebibliography}
%

\end{document}